\documentclass[12pt]{iopart}
\usepackage{amssymb}
\usepackage[mathscr]{eucal}
\usepackage{epic}
\usepackage{eepic}
\usepackage{amsfonts}
\usepackage{bm}

   \newtheorem{theorem}{Theorem}

\newcommand{\nc}{\newcommand}
\nc{\JStP}{{\it J. Stat. Phys.}}  \nc{\IJMP}{{\it Intern. J. Mod. Phys.}}
\nc{\BBS}{{\rm BBS}\ } \nc{\g}{\gamma}  \nc{\lm}{\lambda}
\nc{\la}{\lambda} \nc{\bh}{{\bf h}}  \nc{\av}{\prod_{s\,\in\,\ZN}}
\nc{\cR}{{\cal R}} \nc{\kp}{{\varkappa}} \nc{\om}{\omega}
\nc{\qt}{\tilde{q}} \nc{\tp}{\tilde{p}} \nc{\rt}{\tilde{r}}
\nc{\ty}{\tilde{y}} \nc{\tx}{\tilde{x}} \nc{\tQ}{{\widetilde Q}}
\nc{\trh}{\tilde{\rho}}\nc{\ny}{\nonumber}\nc{\lk}{\left(}
\nc{\rk}{\right)} \nc{\Rb}{\right]} \nc{\Lb}{\left[}
\nc{\rb}{\right\}} \nc{\lb}{\left\{} \nc{\hs}{\hspace*{1cm}}
\nc{\hx}{\hspace*{3mm}} \nc{\hq}{\hspace*{6mm}}

\nc{\al}{\alpha}  \nc{\sig}{\sigma}  \nc{\ZN}{\mathbb{Z}_N}
\nc{\bg}{\boldsymbol{\gamma}} \nc{\bdr}{\boldsymbol{\rho}}
\nc{\bu}{{\bf u}} \nc{\bv}{{\bf v}} \nc{\bV}{{\bf V}}
\nc{\rnk}{r_{n,k}} \nc{\lns}{\la_{n,s}} \nc{\lnl}{\la_{n,l}}
\nc{\FD}{{\cal F}} \nc{\lnk}{\la_{n,k}} \nc{\xnl}{x_{n,l}}
\nc{\Psr}{\Psi_{\bdr_n}} \nc{\ap}{a_{n+1}} \nc{\bp}{b_{n+1}}
\nc{\cp}{c_{n+1}} \nc{\dpp}{d_{n+1}}
\nc{\bep}{\bu_{n+1}^{-1}(\ap-\bp\bv_{n+1})} \nc{\Xop}{\mathbf{X}}
\nc{\Zop}{\mathbf{Z}}  \nc\s{{\gamma}}  \nc{\CD}{{\cal D}}
\def\r#1{(\ref{#1})}
\nc{\sm}[1]{\sum_{#1\in\ZN}}

\nc{\ra}{\rangle} \nc{\BAR}{\begin{array}} \nc{\EAR}{\end{array}}
\nc{\bdm}{\begin{displaymath}} \nc{\edm}{\end{displaymath}}
\nc{\be}{\begin{equation}} \nc{\ee}{\end{equation}}
\nc{\beq}{\begin{equation*}} \nc{\eeq}{\end{equation*}}
\nc{\ba}{\begin{array}} \nc{\ea}{\end{array}}
\nc{\bea}{\begin{eqnarray}} \nc{\eea}{\end{eqnarray}}

\nc\sip{\gamma^\prime}\nc\ma{a'}\nc\mb{b'}\nc\mc{c\,'}\nc\md{d\,'}
\nc\pra{a''}\nc\prb{b''}\nc\prc{c\,''}\nc\prd{d\,''} \nc{\hu}{{\bf
u}}\nc{\hh}{{\hat h}}\nc{\bl}{\boldsymbol{\lambda}}\nc{\hv}{{\bf v}}
\nc\si{{\mathrm{s}}}   \nc{\TS}{{\tilde S}}
\nc{\A}{\mathcal{A}} \nc{\Sc}{\mathcal{S}}  \nc{\N}{\mathcal{N}}
\nc{\rL}{{\rm L}}   \nc{\rR}{{\rm R}}
\nc{\lms}{s}
\nc{\tq}{Q}
\def\qu{{\sf q}}

\def\laxi{\mu}

\begin{document}
\title[Form-factors in the BBS model I: Norms and matrix elements]
   {Form-factors in the Baxter--Bazhanov--Stroganov model I: Norms and matrix elements}
\author{G von Gehlen$^\dag$,~ N Iorgov$^\ddag$,~ S~Pakuliak$^{\sharp\flat}$,  V~Shadura$^\ddag$
 and Yu~Tykhyy$^\ddag$}
\address{$^\dag$\ Physikalisches Institut der Universit\"at Bonn,
Nussallee 12, D-53115 Bonn, Germany}
\address{$^\ddag$ Bogolyubov Institute for Theoretical Physics, Kiev 03680,
Ukraine}
\address{$^\sharp$\ Bogoliubov Laboratory of Theoretical Physics,
Joint Institute for Nuclear Research, Dubna 141980, Moscow region,
Russia}
\address{$^\flat$\ Institute of Theoretical and Experimental Physics,
Moscow 117259, Russia} \ead{gehlen@th.physik.uni-bonn.de,
iorgov@bitp.kiev.ua, pakuliak@theor.jinr.ru,
shadura@bitp.kiev.ua, tykhyy@bitp.kiev.ua}
\begin{abstract}
We continue our investigation of the ${\mathbb Z}_N$-Baxter--Bazhanov--Stroganov model
using the method of separation of variables \cite{gips}. In this paper we calculate the
norms and matrix elements of a local ${\mathbb Z}_N$-spin operator between eigenvectors
of the auxiliary problem.
For the norm the multiple sums over the intermediate states are performed explicitly. In the case
$N=2$ we solve the Baxter equation and obtain form-factors of the spin operator
of the periodic Ising model on a finite lattice.
\end{abstract}
\hspace*{2.5cm}{\small \today}\hspace*{12mm}   \submitto{\JPA}
\vspace*{-12mm} 

\section{Introduction}

The Baxter-Bazhanov-Stroganov model (BBS-model, also known as the
$\tau^{(2)}$-model) \cite{B_tau,BaxInv,BS,BBP} has attracted
considerable interest, because via functional relations it is
related to the solvable $\mathbb{Z}_N$-chiral Potts model (CPM).
Solving these functional relations has been the main method to calculate
the eigenvalues of the CPM \cite{BaxtEV}. However, also by itself, the BBS-model is an
interesting lattice spin model with cyclic $\mathbb{Z}_N$ spin-variables, which is closely
related to the six-vertex model at roots of
unity \cite{Kore,BS,Roan07}. The BBS model in its vertex formulation can be solved
using the Functional Bethe ansatz or Separation of Variables (SoV) method~\cite{Skly1}.
For $N=2$ it is equivalent to a generalized free-fermion
Ising model~\cite{Bugrij}.

Using the formulation of the BBS-model in terms of cyclic
$L$-operators~\cite{Kore,BS,Tarasov}, in \cite{gips} we developed a general
method for the construction of its transfer matrix eigenvectors.
Our approach is an adaptation of the SoV
method of \cite{Skly1,KarLeb2} to $N$-state spin chain
models. In order to find the eigenvectors of the transfer matrix we first obtain the
eigenvectors of a certain auxiliary integrable system. Commuting
integrals of this auxiliary system are generated by the
off-diagonal elements of the monodromy matrix of the BBS model.
Then the eigenvectors of the auxiliary system serve as building blocks for the
eigenvectors of the periodic \BBS model. The multi-variable kernel
which relates the respective eigenvectors can be presented as the product of
{\it single variable} functions (SoV). Each of these functions
satisfies the Baxter equation. In \cite{gips} we showed that the existence of
non-trivial solutions of these equations is equivalent to the well known functional
relations \cite{B_tau,BS,BBP} for the transfer matrix of \BBS model.

The goal of this paper is to use the explicit constructions of
\cite{gips} for calculating matrix elements or form factors of local
operators of the \BBS model. Since the eigenvectors are given in terms of
multiple summations, the main concern of our calculations is to
perform these sums explicitly in order to get factorized
expressions for the form factors and norms. For $N=2$ from the results
of our first paper we calculate spin operator matrix elements
of the generalized Ising model.

We hope that the methods developed in \cite{gips} and in this paper will allow to find
analytical formulas for matrix elements of the CPM.

We would like to mention the paper \cite{Babelon}, where matrix elements
of local operators for the Toda chain are calculated using the
eigenvectors of \cite{KarLeb2}.

This paper is organized as follows: In Section 2 and 3 we recall the results of \cite{gips} for
the right eigenfunctions of $B_n(\la)$,
adding the analogous results for the respective left eigenfunctions, pointing out the differences.
 Then we calculate the overlap of the right and left
eigenfunctions of $B_n(\la)$, i.e. the norms. Here the main result will be that the
sum over intermediate quantum numbers can be performed explicitly. In Section 5 we
concentrate on the case $N=2$ and calculate the overlap of the eigenvectors of the
{\it periodic} transfer matrix $t(\la)=A_n(\la)+D_n(\la)$ for the homogeneous BBS model.
At the end of Section 5 we also write a result for the matrix elements of the single
particle operators ${\bf u}_n$ which still has multiple summations over discrete internal variables.
 In Section 6, using explicit solutions of the Baxter equation
for the case of the homogeneous Ising model, we announce the factorized expression for
the matrix elements of the one site spin operator where all internal summations have been performed.
The details of this calculation will be given in a forthcoming paper \cite{gipst2}.
These will prove a result conjectured by A.~Bugrij and O.~Lisovyy for the matrix elements
of the spin operator in the Ising model on a finite lattice \cite{BL1,BL2}.
Finally, Section 7 summarizes our results and the Appendix explains our summation technique.

\section{$L$-operator formulation of the BBS-model}

We define the BBS-model as a quantum chain model in the $L$-operator vertex formulation
(the relation to the formulation as a face model can be found in \cite{gips,B_tau}).
To each site $k$ of the quantum chain we associate
a cyclic $L$-operator \cite{Kore,BS,Tarasov} acting in a two-dimensional auxiliary space
\be
\label{bazh_strog}L_k(\lm)=\left( \ba{ll}
1+\lm \kp_k \hv_k,\ &  \lm \hu_k^{-1} (a_k-b_k \hv_k)\\ [3mm]
\hu_k (c_k-d_k \hv_k) ,& \lm a_k c_k + \hv_k {b_k d_k}/{\kp_k }
\ea \right)\!\!,\hspace*{4mm} k=1,2,\ldots,n.
\ee
At each site $k$ there are ultra-local Weyl elements $\hu_k$ and
$\hv_k$ obeying the commutation rules and normalization
\beq
\fl \bu_j \bu_k=\bu_k \bu_j\,, \quad\;\bv_j
\bv_k=\bv_k \bv_j\,,\quad\; \bu_j
\bv_k=\om^{\delta_{j,k}}\bv_k\bu_j\,,\quad\; \om=e^{2\pi i/N},
\quad \bu_k^N=\bv_k^N=1\,.
\eeq
$\lm$ is the spectral parameter. We have five parameters
$\:\kp_k,\;a_k,\;b_k,\;c_k,\;d_k\,$ per site. At each site $k$ we
define a $N$-dimensional linear space (quantum space) ${\cal V}_k$
with the basis $ |\g\ra_k$, $\g\in \ZN$,  the dual space ${\cal
V}_k^*$ with the basis $\: _k\langle\g|$, $\g\in \ZN$, and the natural pairing
$\: _k\langle\g'|\g\ra_k=\delta_{\g',\g}.$ In ${\cal V}_k$ the
Weyl elements $\bu_k$ and $\bv_k$ act by the formulas:
\beq
\bu_k |\g\ra_k=\om^\g |\g\ra_k\, ,\qquad \bv_k
|\g\ra_k = |\g+1\ra_k\, .
\eeq
In ${\cal V}_k^*$ these Weyl elements act as follows:
\beq
\: _k\langle\g| \bu_k = \: _k\langle\g| \om^\g\ ,\qquad
\: _k\langle\g| \bv_k  = \: _k\langle\g-1|\, .
\eeq
The monodromy matrix for the quantum chain with $n$ sites is defined by
\be \label{mm}
{T}_n(\lm)\,=\,L_1(\lm)\,L_2(\lm)\,\cdots\, L_n(\lm)= \lk \ba{ll}
A_n(\lm)& B_n(\lm)\\ [.5mm]
C_n(\lm)& D_n(\lm) \ea \rk.
\ee
The transfer-matrix is defined to be the trace in the auxiliary space
$ {\bf t}_{n}(\lambda)\:=\:\mbox{tr}\:
T_n(\lm)\:=\:A_n(\lm)+D_n(\lm)\,.$
This  quantum chain is integrable because the $L$-operators \r{bazh_strog} are
intertwined by the twisted 6-vertex $R$-matrix at roots of unity.
It leads to  $[{\bf t}_{n}(\lambda),{\bf t}_{n}(\mu)]=0$
and so ${\bf t}_n(\la)$ is the generating
 function for the commuting set of non-local and
non-hermitian Hamiltonians ${\bf H}_0,\ldots,{\bf H}_n$ of the model:
\beq
{\bf t}_n(\lm) \:=\:{\bf
H}_0\,+{\bf H}_1\lm\,+\cdots\,+{\bf H}_{n-1}\lm^{n-1}\,+{\bf
H}_{n}\lm^{n}.
\eeq
From the intertwining relation, it follows that $B_n(\lm)$
is the generating function for another commuting set of operators
${\bf h}_1,\ldots,{\bf h}_n$:
\beq
\left[ B_n(\lambda),
B_n(\mu)\right]=0,\qquad B_n(\lm)= {\bf h}_1 \lm + {\bf h}_2 \lm^2
+\cdots + {\bf h}_{n}\lm^{n}\,  .
\eeq
Following Sklyanin \cite{Skly1}, we shall first solve the eigenvalue problem for this last set
of commuting operators, which turns out to be possible by a recursive procedure. After
this, the periodic problem will be related to this auxiliary result by Baxter
equations. After proper normalization by a Sklyanin's measure, the kernel of
the $Q$-operator factorizes into a product of single variable functions (SoV).

\section{Eigenvectors of $B_n(\lm)$}

\subsection{New basis vectors, one-site states}

In the root of unity case an important role is played by
the cyclic function $\:w_p(\g)\:$ \cite{BB} which depends on the
$\ZN$-variable $\g$ and on a point $p=(x,y)$ on the
Fermat curve $\,x^N+y^N=1$. We define $w_p(\g)$ by the difference
equation
\be\frac{w_p(\g)}{w_p(\g-1)}\:=\:\frac{y}{1\,-\,\om^\g\,x}\,,\qquad
\quad w_p(0)=1\,,\qquad \g\in\ZN\,.
\label{Fermat} \ee
The Fermat curve restriction guarantees the cyclic property: $\;w_{p}(\g+N)=w_{p}(\g).$
The function $\:w_p(\g)\:$ is a root of unity analog of the $q$-gamma function.

It is convenient to change the bases in the spaces ${\cal V}_k$
and ${\cal V}_k^*$. Instead of $|\g\ra_k$ and ${}_k\langle \g|$,
$\g\in \ZN$, we will use the vectors
\be\label{psik}\fl  |\psi_{\rho}\ra_k=
\sum_{\g} w_{p^{\rm R}_k}(\g-\rho)|\g\ra_k,\qquad {}_k\langle \psi_{\rho}|= \sum_{\g} \frac{1} {w_{p^{\rm
L}_k}(\g-\rho-1)}\,_k\langle \g| ,\qquad \rho\in \ZN\,.
\ee
The coordinates of the Fermat curve points
$\;p^{\rm L}_k=(x^{\rm L}_k,y^{\rm L}_k)\:$ and $\:p^{\rm R}_k=(x^{\rm
R}_k,y^{\rm R}_k)\:$ are defined as follows.
Let us fix some value of $\,r_k\,$ to satisfy $\:r_k^N=a_k^N-b_k^N$.
(We shall consider generic parameters such that $r_k\ne 0$.)
Then
\beq
x_k=x^{\rm L}_k=x^{\rm
R}_k=\frac{r_k}{a_k},\qquad y_k=y^{\rm R}_k=\frac{b_k}{a_k},
\qquad y^{\rm L}_k=\frac{b_k}{\om a_k}\,.
\eeq
Observe that
$\:y^{\rm R}_k\:=\:\om\,y^{\rm L}_k\,$, while $\:x^{\rm
L}_k\,=\,x^{\rm R}_k$.
The vectors (\ref{psik}) are eigenvectors of the upper off-diagonal matrix
element $\:\lm\bu_k^{-1} (a_k-b_k \bv_k)\,$
of the operator $L_k$:
\be \la\,\bu_k^{-1} (a_k-b_k \bv_k)\, |\psi_{\rho}\ra_k
=\la\,r_k\,\om^{-\rho}\,|\psi_{\rho}\ra_k\,, \label{efbR}
\ee
\be{}_k\langle \psi_{\rho}| \la\,\bu_k^{-1} (a_k-b_k \bv_k)\,
=\la\,r_k\,\om^{-\rho}\,{}_k\langle \psi_{\rho}|\,.\label{efbL}
  \ee
The first equality has been proved in \cite{gips}, the second relation follows very similarly.
The action of the operator $\bv_k$ is
\beq {}_k\langle \psi_{\rho}|
\bv_k= {}_k\langle \psi_{\rho-1}|\,, \qquad \bv_k|\psi_{\rho}\ra_k= |\psi_{\rho+1}\ra_k\,.
\eeq

\subsection{Right and left eigenvectors of $B_n(\lm)$ for general chain length $\;n$}

The eigenvectors of $B_n(\la)$
are labelled by the vector $\bdr_n=(\rho_{n,0},\ldots,\rho_{n,n-1})\in(\ZN)^n$.
Let us further define:
\beq  \trh_n={\textstyle \sum_{k=0}^{n-1}}\;\;\rho_{n,k};\qquad\quad \bdr_n'=
(\rho_{n,1},\ldots,\rho_{n,n-1})\in(\ZN)^{n-1} \eeq
and $\bdr_n^{\pm k}\;$ denotes the vector $\;\bdr_n\;$ in which $\:\rho_{n,k}\:$ is
replaced by $\;\rho_{n,k}\pm 1$, i.e.\\[1mm]
$\hs\hs\hx\;\:\bdr_n^{\pm k}=(\rho_{n,0},\ldots,\rho_{n,k}\pm 1,\ldots,\rho_{n,n-1}),\hq
k=0,1,\ldots,n-1$.\\[-3mm]

The following formulas give an iterative procedure to obtain the eigenvectors
of $\;B_n(\lm)\;$ from eigenvectors of $\:B_{n-1}(\lm)\,$
and single site vectors defined by (\ref{psik}).

The vector \be
|\Psi_{\bdr_n}\ra=\sum_{\bdr_{n-1}\in (\ZN)^{n-1}\atop {\rho_n} \in
\ZN}
 Q^R(\bdr_{n-1},\rho_n|\bdr_n) |\Psi_{\bdr_{n-1}}\ra\otimes
|\psi_{\rho_n}\ra_n \label{PSI}\ee
where \\ [-13mm]
\bea
Q^R(\bdr_{n-1},\rho_n|\bdr_{n})&=&\frac{\om^{(\tilde \rho_n-
\tilde \rho_{n-1}) (\rho_n-\rho_{n,0})}}
{w_{p_{n,0}}(\rho_{n,0}-\rho_{n-1,0}-1) w_{\tilde p_n}(\tilde
\rho_n- \rho_n-1)}\: \times \ny \\ [2mm]  \label{QQQ}
&&\hspace*{-2cm}\times\: \frac{ \prod_{l=1}^{n-2}
\prod_{k=1}^{n-1} w_{p_{n-1,l}^{n,k}}(\rho_{n-1,l}-\rho_{n,k})}
{\prod_{j,l=1\atop j\ne l}^{n-2}
w_{p_{n-1,j}^{n-1,l}}(\rho_{n-1,j}-\rho_{n-1,l})}
\prod_{l=1}^{n-2} \frac{w_{p_{n-1,l}}(-\rho_{n-1,l})}{w_{\tilde
p_{n-1,l}}(\rho_{n-1,l})}\,,\eea
is right eigenvector of $\;B_n(\la)$:\\ [-6mm]
\be\label{Blm}
B_n(\lm)\:|\Psi_{\bdr_n}\ra =\lm\, r_{n,0}\om^{-\rho_{n,0}}
\prod_{k=1}^{n-1}\lk\lm+r_{n,k}\om^{-\rho_{n,k}}\rk|\Psi_{\bdr_n}\ra
\ee
if $|\Psi_{\bdr_{n-1}}\ra$ is right eigenvector of $B_{n-1}(\la)$.
The Fermat curve points $\;\tp_n,\:p_{n,l},\:\tp_{n,l},\:p_{n',l}^{n,k}$
and $r_{n,k}$ entering (\ref{QQQ}), (\ref{Blm})
are related to the parameters of the model $\:a_s,b_s,c_s,d_s,\varkappa_s\:$
by equations (54)--(61) of \cite{gips}, e.g.
$\,\tp_{n,l}=({\tilde x}_{n,l},{\tilde y}_{n,l}),
\;\;{\tilde x}_{n,l}=d_n/(\varkappa_nc_nr_{n,l}),\;\; x^{n,k}_{n-1,l}=r_{n,k}/r_{n-1,l}.\:$
For the more involved
determination of the $r_{n,l}$, see Section 2.3 of~\cite{gips}.\\[-5mm]

Analogously, the
vector \be \langle\Psi_{\bdr_n}|=\sum_{\bdr_{n-1}\in
(\ZN)^{n-1}\atop {\rho_n} \in \ZN}
 Q^{\rm L}(\bdr_{n-1},\rho_n|\bdr_n) \langle\Psi_{\bdr_{n-1}}|\otimes
{}_n\langle\psi_{\rho_n}| \label{PSIL}\ee \vspace*{-4mm} where \\[-8mm]
\bea\fl Q^{\rm L}(\bdr_{n-1},\rho_n|\bdr_{n})&=&
\om^{-(\tilde \rho_n- \tilde \rho_{n-1})
(\rho_n-\rho_{n,0})+\tilde
\rho_{n-1}-\rho_{n-1,0}}\prod_{j,l=1\atop j\ne l}^{n-2}
w_{p_{n-1,j}^{n-1,l}}(\rho_{n-1,j}-\rho_{n-1,l}-1)
     \: \times \ny \\[-1mm]  \label{QQL}\fl && \times\:
\frac
{w_{p_{n,0}}(\rho_{n,0}-\rho_{n-1,0})
w_{\tilde p_n} (\tilde \rho_n- \rho_n)}
{ \prod_{l=1}^{n-2} \prod_{k=1}^{n-1}
w_{p_{n-1,l}^{n,k}}(\rho_{n-1,l}-\rho_{n,k}-1)}
\prod_{l=1}^{n-2}
\frac{w_{\tilde p_{n-1,l}}(\rho_{n-1,l})}{w_{p_{n-1,l}}(-\rho_{n-1,l})}\,,\eea
is left eigenvector of $\;B_n(\la)$:\\ [-6mm]
\beq
\langle\Psi_{\bdr_n}| B_n(\lm) =\lm\,
r_{n,0}\om^{-\rho_{n,0}}
\prod_{k=1}^{n-1}\lk\lm+r_{n,k}\om^{-\rho_{n,k}}\rk \langle\Psi_{\bdr_n}|
\eeq
if $\langle\Psi_{\bdr_{n-1}}|$ is left eigenvector of
$B_{n-1}(\la)$. The definition of $Q^{\rm L}(\bdr_{n-1},\rho_n|\bdr_n)$ uses the same
Fermat curve points
$\;\tp_n,\:p_{n,l},\:\tp_{n,l},\:p_{n',l}^{n,k}$ as were used for the
right eigenvectors. The proof of \r{QQL} by induction is lengthy,
but analogous to the proof of \r{QQQ} given in \cite{gips}.

At $\lm$ equal to one of the $\:n-1\:$ zeros $\:\la_{n,k}\:$ of the
eigenvalue polynomial of $\;B_n(\la)\;$
\be\la_{n,k}\:=\:-r_{n,k}\om^{-\rho_{n,k}},\hs\hs k=1,\ldots,n-1,\label{zeros}\ee
the operators $\:A_n(\lm_{n,k})\:$ and  $\:D_n(\lm_{n,k})\:$ act as shift operators
 for the $k$-th component of the vector ${\bdr_n}$, $\:D_n(\la_{n,k})\:$ shifting in
 addition also the zeroth component:
\be\fl
 A_n(\lm_{n,k})|\Psi_{\bdr_n}\ra\,=\, \varphi_k(\bdr_n')\;|\Psi_{\bdr_n^{+k}}\ra\,,
 \qquad\;\;\,
\langle \Psi_{\bdr_n}|A_n\lk\lm_{n,k}\rk \,=\, \om^{-1}\varphi_k(\bdr_n'^{-k})\;
\langle\Psi_{\bdr_n^{-k}}|\,,\label{Almk}
 \ee
\be\fl
D_n(\lm_{n,k})|\Psi_{\bdr_n}\ra\,=\,\tilde\varphi_k(\bdr_n')\,|\Psi_{\bdr_n^{+0,-k}}\ra ,\qquad
\langle \Psi_{\bdr_n}| D_n(\lm_{n,k})\, =\,\om \,\tilde\varphi_k(\bdr_n'^{+k})\,\langle
\Psi_{\bdr_n^{-0,+k}}|\,, \label{Dlmk}
\ee
 where\\[-11mm]
\bea
\varphi_k(\bdr_n')&=&-\frac{\tilde r_{n-1}}{r_n}\;\om^{-\tilde
\rho_{n}+\rho_{n,0}}\;F_n(\lm_{n,k}/\om)\;\prod_{s=1}^{n-2}y_{n-1,s}^{n,k}\,,
 \ny\\[-2mm]
 \tilde\varphi_k(\bdr'_n)&=&
-\frac{r_n}{\tilde r_{n-1}}\;\frac{\om^{\trh_n-\rho_{n,0}-1}}
{\prod_{s=1}^{n-2}y_{n-1,s}^{n,k}}\;\prod_{m=1}^{n-1}\,F_m(\lm_{n,k})
\eea  and \\[-9mm]
\be F_m(\lm)\:=\:\lk\, b_m\,+\om a_m \,\kp_m \lm\rk\,
\lk \,\la\, c_m\,+d_m/\kp_m\, \rk. \label{qdet}
\ee
The operator $\bV_n=\bv_1\bv_2\ldots\bv_n$, which defines  the term of highest degree in $\lm$ in
$A_n(\la)$ and the free term  in $D_n(\la)$,
shifts the zeroth index of vector ${\bdr_n}$:
\be\label{vvvPsi} \bV_n\,|\Psi_{\bdr_n}\rangle\,=\,|\Psi_{\bdr_n^{+0}}\rangle\,,\qquad
\langle\Psi_{\bdr_n}| \bV_n\,=\,\langle \Psi_{\bdr_n^{-0}}|.
\ee
Using interpolation polynomials and
formulas (\ref{Almk}), (\ref{Dlmk}) and (\ref{vvvPsi}), one can construct how
$A_n(\la)$ and $D_n(\la)$ act on the left eigenvectors of $B_n(\la)$, analogously to what was done
in equations (66)-(68) of \cite{gips} for the right eigenvectors.

\subsection{The norms of the eigenvectors of $B_n(\lm)$}

The pairing $\: _k\langle\g'|\g\ra_k=\delta_{\g',\g}$ implies (we
use $\;y_k\:\equiv\: y_k^R\:=\:\om\, y_k^L$): \be
{}_k\langle\psi_{\rho'}| \psi_\rho\rangle_k =
\delta_{\rho',\rho}\, \frac{N}{\om}\left(\frac{x_k}{y_k}\right)^{N-1}\,. \label{pno}\ee
{\it Proof}.\/ From \r{psik} we get directly \bdm
{}_k\langle\psi_{\rho'}| \psi_\rho\rangle_k =
\sm{\g}\frac{w_{p^{\rm L}_k}(\g)\;\om^\g}{w_{p^{\rm
L}_k}(\g+\rho-\rho'-1)} \edm since $\;\;w_{p^{\rm
R}_k}(\g)\:=\:w_{p^{\rm L}_k}(\g)\:\om^\g\,.$
States for $\rho'\neq \rho$ belong to different eigenvalues in \r{efbR}, \r{efbL}, so
they are ``orthogonal'' (have vanishing pairing). We then use \r{Fermat} and
\bea
\lefteqn{\frac{y^N}{1-\om^\g\,x}\,=\,\sum_{\g'=0}^{N-1}\om^{\g\g'}x^{\g'},\hs\sm{\g}
\frac{1}{1-\om^\g\,x}\,=\,\frac{N}{y^N}\,,}\ny\\
&& \mbox{and for $\;\;1\le \alpha\le N$:}\hs \sm{\g}\frac{\om^{\alpha\,\g}}{1-\om^\g\,x}
\:=\:N\,\frac{x^{N-\alpha}}{y^N}\,.
\hs\hs \Box
 \label{noo}\eea

Next, we calculate the left-right overlap of the eigenstates of
$B_n(\lm)$ for general $n$.
We shall consider only the generic case of parameters such that
all the eigenvalues of $B_n(\lm)$ are different
(in particular, this is important for the action of $A_n(\lm)$ and $D_n(\lm)$
  on the eigenvectors of $B_n(\lm)$).
Therefore $\langle\Psi_{\bdr_n^{L}}|\Psi_{\bdr_n^R}\rangle=0$ if $\bdr_n^L\neq\bdr_n^R$.
The following theorem gives the value of $\langle\Psi_{\bdr_n^{L}}|\Psi_{\bdr_n^R}\rangle$
(``norm'') for
$\;\bdr_n^L=\bdr_n^R=\bdr_n$:

\begin{theorem}
The norms $\langle \Psi_{\bdr_n}|\Psi_{\bdr_n}\rangle$ are independent
of the phase $\rho_{n,0}$ and their dependence
on $\bdr'_n=(\rho_{n,1},\ldots,\rho_{n,n-1})$ is given explicitly as
\be  \fl\langle \Psi_{\bdr_n}|\Psi_{\bdr_n}\rangle \;=\;
\frac{C_n}{\prod_{l<m}(\la_{n,l}-\la_{n,m})}\;=\;
\frac{C_n}{\prod_{l<m}(r_{n,m}\om^{-\rho_{n,m}}-r_{n,l}\om^{-\rho_{n,l}})}\, ,
\label{norm}\ee
where the normalizing factor $C_n$ is independent of $\bdr_n$ and it is defined recursively by
\be \label{recCn}\fl C_n\;=\;C_{n-1}\:{\cal
N}_n\;N^{n-2}\:\frac{\prod_{k<k'}(r_{n,k}^N-r_{n,k'}^N)\;
\prod_{l<l'}(r_{n-1,l'}^N-r_{n-1,l}^N)}{\prod_{k,l}(r_{n,k}^N-r_{n-1,l}^N)}
\prod_{l=1}^{n-2}\,r^{N+1}_{n-1,l}\,,\ee
$C_1$ and $C_2$ are given by
\beq C_1\;=\;\frac{N}{\om}\lk\frac{x_1}{y_1}\rk^{N-1},\hs C_2\;=\;C_1\frac{N^3}{\om}
\lk\frac{x_2}{y_2{\tilde y}_2 y_{2,0}}\rk^{N-1} 
\eeq
and\\[-8mm]
\be\label{calN} {\cal
N}_n\;=\;\frac{N^3}{\om}\lk\frac{x_n}{y_n y_{n,0}{\tilde
y}_n}\rk^{N-1}\lk\:\prod_{l=1}^{n-2}\frac{\prod_{k=1}^{n-1}\;y_{n-1,l}^{n,k}}{\prod_{j,\:j\ne
l}^{n-2}\;y_{n-1,l}^{n-1,j}}\rk.\ee

\end{theorem}

\noindent
{\it Proof}. We shall give the proof by establishing an explicit recursion
$C_{n-1}\,\rightarrow\, C_{n}$. For $n=2$ there is only a single
  zero $\;r_{2,1}\om^{-\rho_{2,1}}$
and the denominator in \r{norm} is unity.
Similarly, for $n=1$ this denominator is unity too as we see from \r{pno}.

From \r{PSI} and \r{PSIL} we get
\be\fl \langle
\Psi_{\bdr_n}|\Psi_{\bdr_n}\rangle
=\!\!\!\!\!\sum_{\bdr_{n-1}\in (\ZN)^{n-1}\atop {\rho_n \in
\ZN}}\!\!\!\!
 Q^{\rm L}(\bdr_{n-1},\rho_n|\bdr_n) \;Q^{\rm R}(\bdr_{n-1},\rho_n|\bdr_n)\;
 \langle \Psi_{\bdr_{n-1}}|\Psi_{\bdr_{n-1}}\rangle
\; {}_n\langle\psi_{\rho_n}|\psi_{\rho_n}\rangle_n .\label{BNO}\ee
Inserting here the explicit expressions \r{QQL} for $Q^{\rm L}$, and  \r{QQQ} for $Q^{\rm R}$,
using \r{pno}, and performing the summations over $\rho_{n-1,0}$ and $\rho_n$ by \r{noo}
we get: \bea \fl
\lefteqn{\langle \Psi_{\bdr_n}|\Psi_{\bdr_n}\rangle \:=\:{\cal
N}_n\sum_{\bdr'_{n-1}\in
(\ZN)^{n-2}}\hspace*{-4mm}\om^{\rho_{n-1,1}+\ldots+\rho_{n-1,n-2}}\;\times}\ny\\
\fl &&\hspace*{7mm}\times\,
\prod_{l=1}^{n-2}\,(r_{n-1,l}\,\om^{-\rho_{n-1,l}})^2\;\frac{\prod_{j,j\neq
l}^{n-2}\lk r_{n-1,j}\,\om^{-\rho_{n-1,j}}-r_{n-1,l}\,\om^{-\rho_{n-1,l}}\rk}
{\prod_{k=1}^{n-1}\lk r_{n-1,l}\,\om^{-\rho_{n-1,l}}-r_{n,k}\,\om^{-\rho_{n,k}}\rk}
\:\langle \Psi_{\bdr_{n-1}}|\Psi_{\bdr_{n-1}}\rangle\,,\label{intNo}\eea
where we combined several phase-independent factors into the
quantity \r{calN}. The main issue now is to
perform the summations over the phases $\bdr'_{n-1}$ explicitly.
In order to avoid too many indices, let us define
\bea
\fl\mathbf{a}^{(n)}\!\!&=&\!\!(a_1,\ldots,a_{n-2})\:=\:\lk
r_{n-1,1}\,
\om^{-\rho_{n-1,1}},\ldots,r_{n-1,n-2}\,\om^{-\rho_{n-1,n-2}}\rk;\ny\\[1mm]\fl
  \mathbf{b}^{(n)}\!\!&=&\!\!(b_1,\ldots,b_{n-1})\:=\:(-\la_{n,1},\ldots,-\la_{n,n-1})\;=\;
  \lk r_{n,1}\om^{-\rho_{n,1}},\ldots, r_{n,n-1}\om^{-\rho_{n,n-1}}\rk. \label{defi}\eea
Then using the induction assumption that
 the formula for the norms is given by \r{norm}, the relation \r{intNo} reduces to
\bea \fl\frac{C_n}{\prod_{l<m}(b_m-b_l)}&=&\N_n\sum_{\rho_{n-1,1},\ldots,\rho_{n-1,n-2}}\;
  \prod_{l=1}^{n-2} r_{n-1,l}\; a_l\:
  \frac{\prod_{j\neq l}^{n-2}(a_j-a_l)}{\prod_{k=1}^{n-1}(a_l-b_k)}
  \;\frac{C_{n-1}}{\prod_{m=l+1}^{n-2}(a_m-a_l)}\ny\\
  \fl&=&C_{n-1}\;\N_n\;{\tilde r}_{n-1}'\; S_n(\mathbf{a}^{(n)},\mathbf{b}^{(n)})\,,
\label{recnormPsi}
\eea
where we define $\;{\tilde r}_{n-1}'=r_{n-1,1}\ldots r_{n-1,n-2}\:$ and

\be S_n(\mathbf{a}^{(n)},\mathbf{b}^{(n)})\;=\!\!\sum_{\rho_{n-1,1},\ldots,
             \rho_{n-1,n-2}\in(\mathbb{Z}_N)^{n-2}}\;\prod_{l=1}^{n-2}\lk
a_l\; \frac{\prod_{j=l+1}^{n-2}\lk
a_l\:-\:a_j\rk}{\prod_{k=1}^{n-1}\lk a_l\:-\:b_k\rk}\rk\,.
\label{SnInt}\ee
In the Appendix, we prove the following identity
\be \fl
 S_n(\mathbf{a}^{(n)},\mathbf{b}^{(n)})\;=\;N^{n-2}
 \lk\prod_{l=1}^{n-2}a^N_l\rk\prod_{k<k'}^{n-1}\frac{b_k^N-b_{k'}^N}{b_k-b_{k'}}\;\:
   \frac{\prod_{l<l'}^{n-2}(a_l^N-a_{l'}^N)}
   {\prod_{l=1}^{n-1}\prod_{l'=1}^{n-2}(b_l^N-a_{l'}^N)}.\label{SnnInt}\ee
The $(\mathbb{Z}_N)^{n-2}$-summation in \r{SnInt} runs over the discrete phases of the variables
$\:\mathbf{a}^{(n)}$. From the $\mathbb{Z}_N$-symmetry the result of this summation
\r{SnnInt} can depend only on the $N$-th powers of these variables.
Substituting \r{SnnInt} into \r{recnormPsi}, and using the notation \r{defi},
 we get the recursion \r{recCn}.
\hfill $\Box$

\subsection{Action of $\bu_n$ on $B_n(\lm)$-eigenvectors}

Now we shall calculate an explicit expression for the
action of $\bu_n$ on the $B_n(\lm)$-eigenvectors $|\Psi_{\bdr_n}\rangle$.
From the relation
\[
\bu_n^{-1} (a_n-b_n \bv_n)|  \psi_{\rho_n}\rangle_n=
r_n \om^{-\rho_n}|  \psi_{\rho_n}\rangle_n\,,
\]
taking into account $a_n/r_n=1/x_n$, $b_n/r_n=y_n/x_n$,  we get
the formula for the action of $\bu_n$ on one-site eigenvectors:
\be\label{unpsi}
\bu_n |  \psi_{\rho_n}\rangle_n=\om^{\rho_n}\left(\frac{1}{x_n} |  \psi_{\rho_n}\rangle_n
- \frac{y_n}{x_n} |  \psi_{\rho_n+1}\rangle_n\right).
\ee
We use this formula to obtain the action of $\bu_n$ on general eigenvectors of $B_n(\lm)$,
which, in particular, will show that most matrix elements of $\bu_n$
between eigenvectors of $B_n(\lm)$ vanish:

\begin{theorem}
The action of $\;\bu_n\,$ on eigenvectors of $\,B_n(\lm)$ is given by
\bea \fl
\lefteqn{\bu_n | \Psi_{\bdr_n}\rangle=\om^{\tilde \rho_n}
 \frac{\tilde x_n}{x_n} | \Psi_{\bdr_n}\rangle-
\om^{\rho_{n,0}}\frac{y_{n,0}y_n}{x_n} | \Psi_{\bdr_n^{+0}}\rangle\;\:+}\label{unPsi}\\
\fl&+&\sum_{k=1}^{n-1}\om^{\rho_{n,0}-\tilde\rho'_n}
\left(\frac{\tilde y_n \tilde r_n}{\om^{-\rho_{n,k}} r_{n,k}r_{n,0}x_n}-
\frac{x_{n,0}\tilde r_{n-1} y_n}{\om \,r_{n-1,0}x_n} \right)
\frac{\prod_{l=1}^{n-2}y^{n,k}_{n-1,l}}{\prod_{s\ne k}
        (-r_{n,k}\om^{-\rho_{n,k}}+r_{n,s}\om^{-\rho_{n,s}})}
|\Psi_{\bdr_n^{+k}}\rangle\,.\ny
\eea
\end{theorem}

\noindent
{\it Proof.}
To prove \r{unPsi} we use \r{PSI}, \r{unpsi} and rewrite \r{unPsi} as a relation for
$Q^{\rm R}(\bdr_{n-1},\rho_n|\bdr_{n})$.
The terms of the resulting relation can be separated into two groups, giving two relations
 which can be proved independently:
\be\label{Qu1} \fl \om^{\rho_n}\, =\,\om^{\tilde \rho_n}\tilde x_n
+\sum_{k=1}^{n-1}\frac{\om^{\rho_{n,0}-\tilde\rho'_n}\tilde y_n
\tilde r_n}{r_{n,k}\om^{-\rho_{n,k}}\:r_{n,0}}\;
\frac{\prod_{l=1}^{n-2}y^{n,k}_{n-1,l}}{\prod\limits_{s\ne k}
(-r_{n,k}\om^{-\rho_{n,k}}+r_{n,s}\om^{-\rho_{n,s}})} \frac{Q^{\rm
R}(\bdr_{n-1},\rho_n|\bdr_{n}^{+k})} {Q^{\rm
R}(\bdr_{n-1},\rho_n|\bdr_{n})} \ee
and \bea \fl\lefteqn{\om^{\rho_n-1} \frac{Q^{\rm
R}(\bdr_{n-1},\rho_n-1|\bdr_{n})} {Q^{\rm
R}(\bdr_{n-1},\rho_n|\bdr_{n})}= \om^{\rho_{n,0}}y_{n,0}
\frac{Q^{\rm R}(\bdr_{n-1},\rho_n|\bdr_{n}^{+0})} {Q^{\rm
R}(\bdr_{n-1},\rho_n|\bdr_{n})}}\ny\\ \fl&&
\hs+\:\sum_{k=1}^{n-1}\om^{\rho_{n,0}-\tilde\rho'_n-1}
\frac{x_{n,0}\tilde r_{n-1}}{r_{n-1,0}}
\frac{\prod_{l=1}^{n-2}y^{n,k}_{n-1,l}}{\prod\limits_{s\ne k}
(-r_{n,k}\om^{-\rho_{n,k}}+r_{n,s}\om^{-\rho_{n,s}})} \frac{Q^{\rm
R}(\bdr_{n-1},\rho_n|\bdr_{n}^{+k})} {Q^{\rm
R}(\bdr_{n-1},\rho_n|\bdr_{n})}\,.\label{Qu2} \eea
In order to verify both \r{Qu1} and \r{Qu2} we start evaluating the ratios
of the $Q^R$ using the explicit formula \r{QQQ}, e.g.
\be\fl
\frac{Q^{\rm R}(\bdr_{n-1},\rho_n|\bdr_{n}^{+k})} {Q^{\rm
R}(\bdr_{n-1},\rho_n|\bdr_{n})}=\om^{\rho_n-\rho_{n,0}}\frac{1-\om^{
\tilde\rho_n-\rho_n}{\tilde x}_n}{\tilde y_n}\prod_{l=1}^{n-2}
\frac{r_{n-1,l}\om^{-\rho_{n-1,l}}-r_{n,k}\om^{-\rho_{n,k}}}
       {r_{n-1,l}\om^{-\rho_{n-1,l}}\:y^{n,k}_{n-1,l}}
\label{Qkshift}\ee
since in \r{Qkshift} the shift $\bdr_{n}^{+k}$ affects only three terms in \r{QQQ} and we use
\r{Fermat} and $x^{n,k}_{n-1,l}=r_{n,k}/r_{n-1,l}.$ For \r{Qu1}, inserting
\r{Qkshift}, and after some cancellations, collecting the $k$-dependent terms, the
sum over $k$ can be performed using the identity
\be \fl\sum_{k=1}^{n-1}\frac{\prod_{l=1}^{n-2}(\xi_k\,-\,\zeta_l)}
  {\xi_k\:\prod_{s=1,s\neq k}^{n-1}(\xi_k\,-\xi_s)}\;=\;\frac{\prod_{l=1}^{n-2}\;\zeta_l}
 {\prod_{k=1}^{n-1}\:\xi_k},\hx \xi_k=-r_{n,k}\om^{-\rho_{n,k}},
 \hx\zeta_l=-r_{n-1,l}\om^{-\rho_{n-1,l}}.\ee
For verifying \r{Qu2} we have to calculate the other shifts of $Q^R$ too.
This time, with the same $\:\xi_k\:$ and $\:\zeta_l\:$ as before,
we perform the sum over $k$ using the identity
\[ \sum_{k=1}^{n-1}\;\frac{\prod_{l=1}^{n-2}(\xi_k\,-\,\zeta_l)}
  {\prod_{s=1,s\neq k}^{n-1}(\xi_k\,-\xi_s)}\;=\;1.
   \hspace*{7.5cm} \Box  \]

\section{Periodic model. Baxter equation}
\label{periodic}

To treat the periodic model
it is convenient to make a Fourier transform over $\rho_{n,0}$
of the eigenvectors of $B_n(\la)$. This yields a basis of eigenvectors
of the operator $\bV_n$ of \r{vvvPsi}:
\beq \langle\tilde\Psi_{\rho,\bdr'_n}|=
{\textstyle \sum_{\rho_{n,0}\in\ZN}}\om^{\rho\cdot\rho_{n,0}}
     \langle\Psi_{\bdr_n}|,\quad
     |\tilde\Psi_{\rho,\bdr'_n}\rangle=
{\textstyle \sum_{\rho_{n,0}\in\ZN}}\om^{-\rho\cdot\rho_{n,0}}
     |\Psi_{\bdr_n}\rangle,
\eeq
\be
\langle\tilde\Psi_{\rho,\bdr'_n}|\bV_n \;=\;\om^\rho\: \langle\tilde\Psi_{\rho,\bdr'_n}|\,,\qquad\qquad
\bV_n |\tilde\Psi_{\rho,\bdr'_n}\rangle \;=\;\om^\rho\: |\tilde\Psi_{\rho,\bdr'_n}\rangle\,.\label{tPsi}
\ee
$\bV_n$ is the total $\mathbb{Z}_N$-spin rotation operator and $\rho\in\mathbb{Z}_N$
is the corresponding total charge.

Let $\langle\Phi_{\rho,{\bf E}}|$ and $|\Phi_{\rho,{\bf E}}\rangle$ be left and right
eigenvectors of ${\bf t}_n(\lm)$ with eigenvalue
\be t_n(\lm|\,\rho,\,{\bf E})\;=\;
E_0+E_1\lm+\cdots+E_{n-1}\lm^{n-1}+E_{n}\lm^{n}\,, \label{tnh}
\ee
where ${\bf E}=\{E_1,\ldots,E_{n-1}\}$ and
the values of $E_0$ and $E_n$
are
\be\label{iE0En}
 E_0\,=1+\om^\rho \prod_{m=1}^n\frac{b_m d_m}{\kp_m},\qquad
E_n\,=\prod_{m=1}^n a_m c_m\,+\om^{\rho}\,\prod_{m=1}^n\kp_m\,.
\ee
Following the general procedure of the SoV method, we express these eigenvectors as
\be \fl\langle\Phi_{\rho,{\bf E}}|\:=\:
\sum_{\bdr'_n} \; Q^{\rm L}(\bdr'_{n}|\,\rho,{\bf E})\;\langle\tilde\Psi_{\rho,\bdr'_n}|\,,\qquad
|\Phi_{\rho,{\bf E}}\rangle \:=\:
\sum_{\bdr'_n} \; Q^{\rm R}(\bdr'_{n}|\,\rho,{\bf E})|\tilde\Psi_{\rho,\bdr'_n}\rangle\,, \label{EV}
\ee
where\footnote{The kernels in \r{EV}, \r{QL} and \r{QR}
are not the kernels \r{QQQ} and \r{QQL} used in the recursive definitions of the left
and right eigenvectors of $B_n(\lm)$, observe the different types of arguments.}\\[-12mm]
\be \label{QL}
Q^{\rm L} (\bdr'_{n}|\,\rho,{\bf E})=\prod_{k=1}^{n-1}\:
\tq_k^{\rm L}(\rho_{n,k}) \prod_{s,s'=1\atop s\ne s'}^{n-1}
w_{p_{n,s}^{n,s'}}(\rho_{n,s}-\rho_{n,s'}-1),
\ee
\be \label{QR}
Q^{\rm R} (\bdr'_{n}|\,\rho,{\bf E})=\frac{\prod_{k=1}^{n-1}\:
\tq_k^{\rm R}(\rho_{n,k})}{ \prod_{s,s'=1\atop s\ne s'}^{n-1}
w_{p_{n,s}^{n,s'}}(\rho_{n,s}-\rho_{n,s'})}\,.
\ee
The products of $\ZN$-cyclic functions $w_{p_{n,s}^{n,s'}}(\rho_{n,s}-\rho_{n,s'})$ in
\r{QL} and \r{QR} are Sklyanin's measure which makes the rest of the kernels
$\:Q^{\rm L} (\bdr'_{n}|\,\rho,{\bf E})$ and
$\:Q^{\rm R} (\bdr'_{n}|\,\rho,{\bf E})$ factorizable into products of single variable functions (SoV).
The Baxter equations for these functions of separated variables are
\be\fl
t_n(\lm_{n,k}|\rho,{\bf E})\;\tq^{\rm L}_k(\rho_{n,k})=
\om^{n-1}\Delta_k^-(\lm_{n,k})\;\tq^{\rm L}_k(\rho_{n,k}+1)+\om^{1-n}\Delta_k^+(\om\lm_{n,k})\;
\tq^{\rm L}_k(\rho_{n,k}-1)\,,\label{BAXL}\ee
\be\fl
t_n(\lm_{n,k}|\rho,{\bf E})\;\tq^{\rm R}_k(\rho_{n,k})=
\Delta_k^+(\lm_{n,k})\;\tq^{\rm R}_k(\rho_{n,k}+1)+\Delta_k^-(\om\lm_{n,k})\;
\tq^{\rm R}_k(\rho_{n,k}-1)\,,\label{BAXR}\ee
where $\:\la_{n,k}=-r_{n,k}\om^{-\rho_{n,k}},\;\;k=1,\ldots,n-1,\:$ are the zeros of the
eigenvalue polynomial of $B_n(\la)$, see \r{zeros}, and
\beq \fl\Delta_k^+(\lm)\:=\:(\om^\rho/\chi_k)\,(\lm/\om)^{1-n}\:
         \prod_{m=1}^{n-1}\,F_m(\lm/\om)\,,\qquad
\Delta_k^-(\lm)\:=\:\chi_k\:(\lm/\om)^{n-1}\:F_n(\lm/\om)\,, \eeq
\be
\chi_k\;=\;\frac{r_{n,0}\:\rt_{n-1}}{r_n\:\rt_n}\;\:(\prod_{s=1\atop s\ne
k}^{n-1}\; y_{n,k}^{n,s}/y^{n,k}_{n,s})\:\prod_{s=1}^{n-2}\:y^{n,k}_{n-1,s}\,. \label{Dpm}
\ee

\section{Periodic \BBS model for $N=2$}
\label{perN2}

\subsection{Eigenvalues of transfer matrix for generalized homogeneous Ising model}

In this section we consider in more detail
 the case of the $N=2$ periodic homogeneous \BBS model, where $\;\om=-1$.
By homogenous we mean that the parameters $a$, $b$, $c$, $d$ and $\kp$ each are taken to be
independent of the site index. As it was shown in \cite{Bugrij}
this model is a particular case (``free
fermion point'') of the generalized Ising model.

In the  $N=2$ case, we have $\hu_k\,=\:\sigma^z_k$ and $\hv_k\,=\:\sigma^x_k$, where
$\sigma^z_k$ and $\sigma^x_k$ are Pauli matrices acting on the spin at the $k$-th site.
The operator $\bV_n=\sigma^x_1 \sigma^x_2 \cdots \sigma^x_n$ is the spin-flip operator.
It commutes with the transfer-matrix ${\bf t}_n(\lm)$, and we have simply $\bV_n^2=1$.
Therefore the eigenvectors of ${\bf t}_n(\lm)$ are divided into two sectors according to the
eigenvalue $\:(-1)^\rho$, $\rho\in\{0,1\}$ of $\bV_n$. If the parameters $a$, $b$, $c$, $d$ and $\kp$
are generic, no degeneracies of the eigenvalues occur. The transfer-matrix of the
 standard Ising model on the finite lattice can be obtained from the $N=2$ BBS integrals
 of motion ${\bf H}_k$, $k=0,1,\ldots,n$,  when
parameters of the homogeneous $N=2$ BBS model satisfy the relations\\[-9mm]
\be a=c,\hq d=b,\hq \kp=1\hq \mbox{and}\hq \lm=b/a\label{spIsi}
\ee
(see Section~\ref{sect61}). In this case degenerations of the
spectra occur and in order to distinguish the corresponding eigenvectors one has to use an additional
operator which commutes with transfer matrix. In case of the periodic model, such an operator
is the operator of translation by one site:
\[
{\bf T}\; |\gamma_1\rangle_1\otimes |\gamma_2\rangle_2\otimes\cdots \otimes|\gamma_n\rangle_n=
|\gamma_n\rangle_1\otimes |\gamma_1\rangle_2\otimes\cdots \otimes|\gamma_{n-1}\rangle_n\,,
\]\[
{\bf T}\; \sigma^{z,x}_k {\bf T}^{-1}=\sigma^{z,x}_{k+1},\qquad
[\,{\bf T},{\bf t}_n(\lm)]=0, \qquad
[\,{\bf T},\bV_n]=0\,.
\]
Our construction of the eigenvectors  \r{EV} is obviously non-invariant with respect
to translations and we were not able to show that the vectors \r{EV} are  also eigenvectors of ${\bf T}$.
A translational invariant description of the eigenvectors for the
generalized Ising model has been given in \cite{Lisovyy}. The description of the spectra of
the transfer matrix
given in this paper coincides with the description of the eigenvalues obtained in our
formalism presented in \cite{gips}. This allows to identify the eigenvectors in
both descriptions. In particular, the
eigenvectors are labelled by quasi-momenta of the excitations $\:\qu=\pm\frac{2\pi}{n}m\,$, where $m$
is integer or half-integer depending the eigenvalue $(-1)^\rho$ of $\,\bV_n$.
So the whole space of states decomposes into two
sectors according to the value of $\rho$:
\begin{itemize}
\item
NS--sector: $\rho=0$, the eigenstates of ${\bf t}_n$ have an {\it even} number of excitations with
 quasi-momenta
$\;\qu\in {\rm NS} \equiv \left\{\frac{2\pi}n (\mathbb{Z}_n+1/2)\right\}$. The dimension of this
sector is $2^{n-1}$.
\item
R--sector: $\rho=1$, the eigenstates of ${\bf t}_n$ have an {\it odd} number
of excitations with quasi-momenta
$\;\qu\in {\rm R} \equiv \left\{\frac{2\pi}n \mathbb{Z}_n\right\}$.
The dimension of this sector is also $2^{n-1}$.
\end{itemize}
In \cite{gips} the eigenvalues $t(\la)$ (we suppress the chain-length index $n$)
were derived from a functional equation which could be written in the form
(don't confuse these $A(\qu)$ etc. with the $A_n(\la)$ etc. in \r{mm}):
\be  t(\la)\,t(-\la)\,=\;(-1)^n\prod_\qu\lk A(\qu)\la^2\,-C(\qu)\,+2i\la\,B(\qu)\rk, \label{funceq}\ee
where\\[-11mm] \be \ba{l}  A(\qu)\,=\,a^2\,c^2\,-2\kp\, ac\,\cos \qu\,+\kp^2,\hs
B(\qu)\,=\,(a\,d\,-b\,c)\sin \qu\,,\ny\\[2mm]
C(\qu)=1\,-2(b\,d/\kp)\cos \qu\, + b^2d^2/\kp^2\,.\ea\label{C}\ee
Solving \r{funceq}, the $2^n$ eigenvalues corresponding to the NS- and R-sectors are given by \r{tNSR}
\cite{gips,Bugrij,Lisovyy} where we have to choose all possible sets of
$\:\pm-$signs ~(which we write $(-1)^{\sig_\qu}$
with $\:\sig_\qu\in\{0,1\}$):
\be
\begin{array}{lrcll}\fl
{\rm NS}: &t(\lm)&=&(a^n c^n+\kp^n)\prod_{\qu\in {\rm NS}} (\lm\,+\,(-1)^{\sigma_\qu}\lms_\qu),\hq&
\prod_{\qu\in {\rm NS}}(-1)^{\sig_\qu}\,=\,+1,\ny\\[2mm] \fl
{\rm R}:\;\;\, &t(\lm)&=&(a^n c^n-\kp^n)\:\prod_{\qu\in {\rm R}}\, (\lm\,+\,(-1)^{\sigma_\qu}\lms_\qu),\hq \;
&\prod_{\qu\in {\rm R}}\:(-1)^{\sig_\qu}\,=\,-1,
\end{array}
\label{tNSR}
\ee
the restriction on the signs follows from \r{tPsi}.
We shall also write
$\; \la_\qu\;=\;(-1)^{\sig_\qu}\,s_\qu, \;$ and
the amplitudes $\lms_\qu$ are given by
\bea  \lms_\qu\;=\;(\sqrt{D(\qu)}\,-{\rm i} B(\qu))/A(\qu),\qquad
D(\qu)\;=\;A(\qu)\,C(\qu)\,-B(\qu)^2.        \label{57}\eea
The convention which fixes the sign of $\sqrt{D(\qu)}$ is given in (120),(121) of \cite{gips}.
States which have eigenvalues where in \r{tNSR} there are some minus signs
$(-1)^{\sig_\qu}=-1$ are said to have excitations with quasi-momenta $\qu$.
The restriction on the signs means that for the NS (R) sector the number of excitations,
i.e. the number of minus signs, must be even (odd).

As an example, the $2^3$ eigenstates and eigenvalues for the $3$-site chain ($n=3$) are:
\\[-6mm]
\begin{itemize}
\item
NS--sector $(\rho=0)$:\\[-6mm]
\[\fl
\ba{lll} |\rangle_{\rm NS} \mbox{\ (no excitations)}  &\to&
t(\lm)=(a^3c^3+\kp^3)(\lm+\lms_{\pi/3})(\lm+\lms_{-\pi/3})(\lm+\lms_\pi),\\
|\pm\pi/3,\pi\rangle_{\rm NS},\quad
&\to& t(\lm)=(a^3c^3+\kp^3)(\lm\mp \lms_{\pi/3})(\lm\pm\lms_{-\pi/3})(\lm-\lms_\pi),\\
|\pi/3,-\pi/3\rangle_{\rm NS}, \quad &\to& t(\lm)=(a^3c^3+\kp^3)(\lm-\lms_{\pi/3})
(\lm-\lms_{-\pi/3})(\lm+\lms_\pi).
\ea
\]
\item
R--sector $(\rho=1$):\\[-6mm]
\[\fl
\ba{lll} |0\rangle_{\rm R} &\to&
t(\lm)=(a^3c^3-\kp^3)(\lm+\lms_{2\pi/3})(\lm+\lms_{-2\pi/3})(\lm-\lms_0),\\
|\pm 2\pi/3\rangle_{\rm R}\quad &\to&t(\lm)=(a^3c^3-\kp^3)(\lm\mp\lms_{2\pi/3})
(\lm\pm\lms_{-2\pi/3})(\lm+\lms_0),\\
|0,2\pi/3,-2\pi/3\rangle_{\rm R} \quad &\to& t(\lm)=(a^3c^3-\kp^3)(\lm-\lms_{2\pi/3})
(\lm-\lms_{-2\pi/3})(\lm-\lms_0).
\ea\]\end{itemize}

From \cite{Lisovyy}, it follows that
the action of operator of translation ${\bf T}$ on the eigenstates gives eigenvalues
of the form $e^{{\rm i}P}$,
 where $P$ is
the sum of the quasi-momenta of all excitations of the given state.
In case of the Ising model when $ad=bc$ and $B(q)\equiv0$ we have $\lms_\qu=\lms_{-\qu}$:
two states which have
excitations with quasi-momenta $\qu$ and $-\qu$ have the same eigenvalues but can be
distinguished by the eigenvalue of
$\:{\bf T}$.

The eigenvalues of ${\bf t}_n(\lm)$ provide the existence of nontrivial solutions of the systems
\r{BAXL}, \r{BAXR} of homogeneous Baxter equations. These solutions give explicit formulas (\ref{EV}) for
the eigenvectors, which form a basis in the space of states of the
periodic \BBS model for $N=2$.

\subsection{Norms and orthogonality of the eigenvectors of the periodic $N=2$ BBS model}

Let us fix an eigenvalue polynomial $t(\lm)$ of $A_n(\lm)+D_n(\lm)$ corresponding to a
right eigenvector from the sector $\rho$.
In order to find this eigenvector explicitly we have to solve
the associated $n-1$ systems ($k=1,2,\ldots,n-1$) of Baxter equations:
\be\ba{rcl}
t(-r_{n,k}) {\tq}^{\rm R}_k(0)&=&
\left(\Delta_k^+(-r_{n,k})+\Delta_k^-(r_{n,k})\right){\tq}^{\rm R}_k(1),
\ny\\[2mm]
t(r_{n,k})\;\; {\tq}^{\rm R}_k(1)&=&
\left(\Delta_k^+(r_{n,k})+\Delta_k^-(-r_{n,k})\right){\tq}^{\rm R}_k(0).
\ea
\label{BEq}
\ee
Since $t(\lm)$ is eigenvalue polynomial, it satisfies the functional relation.
In \cite{gips} it was shown, that the functional relation \r{funceq} ensures the existence
of non-trivial solutions
to \r{BEq} with respect to the unknown variables
${\tq}^{\rm R}_k(0)$ and ${\tq}^{\rm R}_k(1)$ for every $k=1,2,\ldots,n-1$.
In the $N=2$ case, this means that for every $k$ we have one (in the case of degenerate eigenvalues,
possibly zero)
independent linear equation.
In the case of generic parameters, both hand sides of each equation are non-zero. So we may fix
${\tq}^{\rm R}_k(0)=1$\footnote{When the parameters
satisfy the Ising model restrictions,
it is not always possible to choose such normalization.} and
obtain two equivalent expressions for ${\tq}^{\rm R}_k(1)$:
\[
{\tq}^{\rm R}_k(1)=\frac{t(-r_{n,k})}{\Delta_k^+(-r_{n,k})+\Delta_k^-(r_{n,k})}=
\frac{\Delta_k^+(r_{n,k})+\Delta_k^-(-r_{n,k})}{t(r_{n,k})}\,.
\]
Similarly, for the left eigenvector we have
\be\ba{rcl}
t(-r_{n,k})\, {\tq}^{\rm L}_k(0)&=&(-1)^{n-1}\left(\Delta_k^+(r_{n,k})+\Delta_k^-(-r_{n,k})\right)
{\tq}^{\rm L}_k(1)\,,
\ny\\[2mm]
t(r_{n,k})\, {\tq}^{\rm L}_k(1)&=&(-1)^{n-1}
\left(\Delta_k^+(-r_{n,k})+\Delta_k^-(r_{n,k})\right)
{\tq}^{\rm L}_k(0)\,.
\ea\label{LBaxter}\ee
Fixing ${\tq}^{\rm L}_k(0)=1$ we obtain
\[
{\tq}^{\rm L}_k(1)=\frac{(-1)^{n-1} t(-r_{n,k})}{\Delta_k^+(r_{n,k})+\Delta_k^-(-r_{n,k})}=
\frac{\Delta_k^+(-r_{n,k})+\Delta_k^-(r_{n,k})}{(-1)^{n-1}t(r_{n,k})}\,.
\]
Since for generic parameter values we shall
have $t(r_{n,k})\neq 0$, these explicit formulas give immediately
\[
{\tq}^{\rm L}_k(\rho_{n,k}){\tq}^{\rm R}_k(\rho_{n,k})=
(-1)^{\rho_{n,k} (n-1)} t((-1)^{\rho_{n,k}} r_{n,k})/t(r_{n,k})\,.
\]

Eigenvectors in the periodic case are defined by the formula \r{EV}.
In the case $N=2$ we have a simple inversion relation for the cyclic function $w_p(\rho)$:
$w_p(\rho)w_p(\rho-1)=y/(1+x)$. This allows us to use instead of (\ref{QL}) the following kernel
\be \label{QLnew}
Q^{\rm L} (\bdr'_{n}|\,\rho,{\bf E})=\frac{\prod_{k=1}^{n-1}\: \tq_k^{\rm L}(\rho_{n,k})}
{\prod_{s,s'=1\atop s\ne s'}^{n-1} w_{p_{n,s}^{n,s'}}(\rho_{n,s}-\rho_{n,s'})}\,.
\ee
Up to a coefficient, \r{QLnew} with \r{EV} gives
the same eigenvectors of $\:A_n(\lm)+D_n(\lm)\:$ as \r{QL}. In what follows we shall use only \r{QLnew}.
The pairing of $\langle \Phi|$ and $|\Phi\rangle$, which are left and right eigenvectors corresponding to
the eigenvalue $t(\lm)$, gives (we will use the normalization by
$\langle \tilde\Psi_{0,{\bf 0}}|\tilde\Psi_{0,{\bf 0}}\rangle$
to get rid of unimportant factors appearing in the formula for the norms)
\[\fl
\frac{\langle \Phi|\Phi\rangle}{\langle \tilde\Psi_{0,{\bf 0}}|\tilde\Psi_{0,{\bf 0}}\rangle}
=\sum_{\bdr'_n}
\frac{\prod_{k=1}^{n-1}{\tq}^{\rm L}_k(\rho_{n,k}){\tq}^{\rm R}_k(\rho_{n,k})}
{\prod_{l<m}^{n-1} (w_{p_{n,l}^{n,m}}(\rho_{n,l}-\rho_{n,m}) w_{p_{n,m}^{n,l}}
(\rho_{n,m}-\rho_{n,l}))^2}
\frac{\langle \tilde\Psi_{\rho,\bdr'_n}|\tilde\Psi_{\rho,\bdr'_n}\rangle}
{\langle \tilde\Psi_{0,{\bf 0}}|\tilde\Psi_{0,{\bf 0}}\rangle}\,.
\]
where for the norm we have used (\ref{norm}) which leads to
\be
\frac{\langle \tilde\Psi_{\rho,\bdr'_n}|\tilde\Psi_{\rho,\bdr'_n}\rangle}
{\langle \tilde\Psi_{0,{\bf 0}}|\tilde\Psi_{0,{\bf 0}}\rangle}=
\frac{\prod_{l<m}^{n-1} (r_{n,m} (-1)^{\rho_{n,m}}+r_{n,l}(-1)^{\rho_{n,l}})}
{\prod_{l<m}^{n-1} (r_{n,m}+r_{n,l})}\,,
\ee
where $\;{\bf 0}=(0,0,\ldots,0)$ has $n-1$ components.
In the $N=2$ case we have
\[
w_p(1)=\frac{y}{1+x}=\frac{1-x}{y}\,,\hs \frac{1}{(w_p(1))^2}=\frac{1+x}{1-x}\,.
\]
Therefore for the Fermat point $p_{n,l}^{n,m}=(x_{n,l}^{n,m},y_{n,l}^{n,m})$ with coordinate
$\:x_{n,l}^{n,m}=r_{n,m}/r_{n,l}\:$ we have\\[-8mm]
\[
\lk w_{p_{n,l}^{n,m}}(\rho_{n,l}-\rho_{n,m})\rk^{-2}\:=\:
(-1)^{\rho_{n,l}}\frac{(r_{n,l}+r_{n,m})}{(-1)^{\rho_{n,l}} r_{n,l}+ (-1)^{\rho_{n,m}}r_{n,m}}
\]
and so
\[\fl
\frac{1}{(w_{p_{n,l}^{n,m}}(\rho_{n,l}-\rho_{n,m}) w_{p_{n,m}^{n,l}}(\rho_{n,m}-\rho_{n,l}))^2}=
(-1)^{\rho_{n,l}+\rho_{n,m}}\frac{(r_{n,l}+r_{n,m})^2}{((-1)^{\rho_{n,l}} r_{n,l}+
(-1)^{\rho_{n,m}}r_{n,m})^2}\,.
\]
Combining all these formulas we get for the left-right overlap
of the transfer matrix eigenvectors of the periodic BBS model at $N=2$:
\be\fl
\frac{\langle \Phi\,|\,\Phi\rangle}{\langle \tilde\Psi_{0,{\bf 0}}|\tilde\Psi_{0,{\bf 0}}\rangle}\;=\;
\frac{{\prod_{l<m}^{n-1} (r_{n,m}+r_{n,l})}}{\prod_{l=1}^{n-1}\;t(r_{n,l})}
\;\sum_{\bdr'_n}\; \frac{\prod_{l=1}^{n-1} (-1)^{\rho_{n,l}}\:t((-1)^{\rho_{n,l}}r_{n,l})}
{\prod_{l<m}^{n-1} ((-1)^{\rho_{n,m}}r_{n,m}\,+(-1)^{\rho_{n,l}}r_{n,l})}.\label{phiphi}\ee

Using the same techniques as we used for calculating the norm in the auxiliary problem,
we are now able to perform the summations in \r{phiphi} explicitly:
We write the polynomial $t(\la)$, equations \r{tnh}, \r{tNSR}, as
\be t(\la)\;=\;\Lambda\:\prod_{k=1}^n \;(\la\:+\:\laxi_k)\,, \ee
where $\Lambda=a^n c^n+(-1)^\rho\kp^n$ and $\;\laxi_k=\la_\qu$ with
$\;\qu=\frac{2\pi}{n}(k+\frac{1-\rho}{2})$.
Then the sum in \r{phiphi} is
\bea \fl\lefteqn{\Lambda^{n-1}\:\sum_{\bdr'_n}\; \prod_{l=1}^{n-1}\:(-)^{\rho_{n,l}}
   \frac{\prod_{k=1}^n((-)^{\rho_{n,l}}r_{n,l}\,+\,\laxi_k)}
   {\prod_{j=l+1}^{n-1} ((-)^{\rho_{n,j}}r_{n,j}\, +(-)^{\rho_{n,l}}r_{n,l})}}\ny\\
\fl\;\;\;\; &=&\frac{\Lambda^{n-1}}{\tilde r_n'}\prod_{l=1}^{n-1}
\frac{\prod_{k=1}^n(r_{n,l}^2\,-\,\laxi_k^2)}{\prod_{m=l+1}^{n-1} (r_{n,m}^2\,-\,r_{n,l}^2)}
    \sum_{\bdr'_n} \prod_{l=1}^{n-1}(-)^{\rho_{n,l}}r_{n,l}
    \frac{\prod_{j=l+1}^{n-1} ((-)^{\rho_{n,j}}r_{n,j}\,-(-)^{\rho_{n,l}}r_{n,l})}
     {\prod_{k=1}^n(r_{n,l}(-)^{\rho_{n,l}}\,-\,\laxi_k)}\ny\\
\fl &=&\Lambda^{n-1}\prod_{l=1}^{n-1}\frac{\prod_{k=1}^n(r_{n,l}^2\,-\,\laxi_k^2)}{\prod_{m=l+1}^{n-1}
(r_{n,m}^2\,-\,r_{n,l}^2)}\,
  2^{n-1}\,\tilde r_n'\prod_{k<k'}\frac{\laxi_k^2\,-\,\laxi_{k'}^2}{\laxi_k\,-\,\laxi_{k'}}
\frac{\prod_{l<l'}^{n-1}(r_{n,l}^2\,-\,r_{n,l'}^2)}{\prod_{l=1}^n\prod_{l'=1}^{n-1}(\laxi_l^2-
r_{n,l'}^2)}\ny\\
\fl &=& \Lambda^{n-1}\,2^{n-1}\,\tilde r_n'\prod_{k<k'}(\laxi_k\,+\,\laxi_{k'})\,,\hs
\hq\tilde r_n'\,=\,r_{n,1}r_{n,2}\ldots r_{n,n-1}.
  \eea
In the second line the sum takes just the form \r{Sn}
which we had for $S_{n+1}$ with $N=2$ and
$\mathbf{a}^{(n+1)}=(r_{n,1} (-)^{\rho_{n,1}},\ldots,r_{n,n-1} (-)^{\rho_{n,n-1}})$ and
$\mathbf{b}^{(n+1)}=(\laxi_1,\ldots,\laxi_{n})$.
In the last two lines we have used the result  \r{Snn}
for $S_{n+1}$ derived in the Appendix.
 So we obtain the norm of the general periodic state vector
\be\label{normP}
\frac{\langle \Phi\,|\,\Phi\rangle}{\langle \tilde\Psi_{0,{\bf 0}}|\tilde\Psi_{0,{\bf 0}}\rangle}\;=\;
  2^{n-1}\,\tilde r_n'\;
  \frac{{\prod_{l<m}^{n-1} (r_{n,m}+r_{n,l})}}{\prod_{k=1}^{n}\prod_{l=1}^{n-1} (r_{n,l}\,+\,\laxi_k)}
\;\prod_{i<j}^n (\laxi_i\,+\,\laxi_j)\,.
\ee

If the spectrum of the model is not degenerate, the overlap of the eigenvectors corresponding
to different eigenvalues is zero. In case of degeneration, as it
happens in the case of the parameterization $a=c$, $b=d$ and $\kp=1$, corresponding to the Ising model,
one should be careful using \r{normP}. In Section~\ref{sect6}
this case will be discussed in detail.

\subsection{Matrix elements between eigenvectors of the periodic $N=2$ BBS model}

In Theorem 2 we found the action of $\bu_n$ on an eigenvector
$|\Psi_{\bdr_n}\rangle$ of $B_n(\la)$: the result is a linear combination of the original vector
 plus a sum of vectors which each have one component of $\bdr_n$ raised. In order
 to get the matrix elements of $\bu_n$ in the periodic model, we first obtain
 the matrix elements between Fourier transformed eigenvectors of $B_n(\lm)$ defined in \r{tPsi}
\[\fl
\langle \tilde\Psi_{\rho,\bdr'_n}|=\langle \Psi_{0,\bdr'_n}| + (-)^\rho \langle \Psi_{1,\bdr'_n}|\,,\qquad
|\tilde\Psi_{\rho,\bdr'_n}\rangle= | \Psi_{0,\bdr'_n}\rangle + (-)^\rho |\Psi_{1,\bdr'_n}\rangle\,.
\]
Using \r{unPsi}, the following $n$-site matrix elements of
$\bu_n$ are non-zero
\[\fl
\frac{\langle \tilde\Psi_{1,\bdr'_n}|\bu_n|\tilde\Psi_{0,\bdr'_n}\rangle}
{\langle \tilde\Psi_{0,\bdr'_n}|\tilde\Psi_{0,\bdr'_n}\rangle}=
\frac{\tilde x_n}{x_n}(-1)^{\tilde  \rho'_n}+\frac{y_{n,0} y_n}{x_n}=
\frac{a_n}{\tilde r_n}(-1)^{\tilde  \rho'_n}+\frac{\kp_1\kp_2\cdots\kp_{n-1} b_n}{r_{n,0}}\,,
\]\[\fl
\frac{\langle \tilde\Psi_{0,\bdr'_n}|\bu_n|\tilde\Psi_{1,\bdr'_n}\rangle}
{\langle \tilde\Psi_{0,\bdr'_n}|\tilde\Psi_{0,\bdr'_n}\rangle}=
\frac{\tilde x_n}{x_n}(-1)^{\tilde  \rho'_n}-\frac{y_{n,0} y_n}{x_n}=
\frac{a_n}{\tilde r_n}(-1)^{\tilde  \rho'_n}-\frac{\kp_1\kp_2\cdots\kp_{n-1} b_n}{r_{n,0}}\,,
\]\[\fl
\frac{\langle \tilde\Psi_{1,{\bdr'}^{+k}_n}|\bu_n|\tilde\Psi_{0,\bdr'_n}\rangle}
{\langle \tilde\Psi_{0,\bdr'_n}|\tilde\Psi_{0,\bdr'_n}\rangle}=
\frac{\langle \tilde\Psi_{0,{\bdr'}^{+k}_n}|\bu_n|\tilde\Psi_{1,\bdr'_n}\rangle}
{\langle \tilde\Psi_{0,\bdr'_n}|\tilde\Psi_{0,\bdr'_n}\rangle}=
\]
\be\fl\hs
=(-1)^{\tilde  \rho'_n}\frac{\tilde r_{n-1} a_n b_n c_n}{r_n r_{n,0}}
\left(1+\frac{(-1)^{\rho_{n,k}} d_n}{\kp_n c_n r_{n,k}}\right)
\frac{\prod_{l=1}^{n-2}y^{n,k}_{n-1,l}}
{\prod_{s\ne k} (r_{n,k}(-1)^{\rho_{n,k}}+r_{n,s}(-1)^{\rho_{n,s}})}\,,
\label{psiupsi}\ee  where $\tilde \rho'_n=\rho_{n,1}+\rho_{n,2}+\cdots+\rho_{n,n-1}$.
Note, since $\bu_n$ anti-commutes with $\bV_n$, all
matrix elements of $\bu_n$ between the vectors from the same sector $\rho$ are zero.

Then by \r{EV} we pass to periodic eigenstates transforming by the solutions $Q$ of the Baxter equations.
Let $\langle \Phi_0|$ be a left eigenvector  of the transfer-matrix ${\bf t}_n(\lm)$ with $\rho=0$
and $|\Phi_1\rangle$ be a right eigenvector with $\rho=1$.
Let $\tq^{L(0)}_k(\rho_{n,k})$ and $\tq^{R(1)}_k(\rho_{n,k})$
be the solutions of Baxter equation corresponding to these two eigenvectors.
Then for the matrix element after some simplification we have
\bea
\fl\lefteqn{\frac{\langle \Phi_0|\bu_n|\Phi_1\rangle}
{\langle \tilde\Psi_{0,{\bf 0}}|\tilde\Psi_{0,{\bf 0}}\rangle}
\;=\;\sum_{\bdr'_n} (-1)^{n\tilde \rho'_n} \prod_{l<m}^{n-1}
\frac{r_{n,l}+r_{n,m}}{(-1)^{\rho_{n,l}} r_{n,l}+ (-1)^{\rho_{n,m}}r_{n,m}}\:\times}
\ny\\ \fl
\hx&&\times
\left(\:\prod_{l=1}^{n-1}\;{\tq}^{L(0)}_l(\rho_{n,l}){\tq}^{R(1)}_l(\rho_{n,l})
\left(\frac{a_n}{\tilde r_n}(-1)^{\tilde  \rho'_n}-\frac{\kp_1\kp_2\cdots\kp_{n-1} b_n}{r_{n,0}}\right)+
\right.
\ny\\ \fl
 && \hx\hx+\;\sum_{k=1}^{n-1}\:
\tq^{L(0)}_k(\rho_{n,k}+1){\tq}^{R(1)}_k(\rho_{n,k})
\prod_{l\ne k}^{n-1} \tq^{L(0)}_l(\rho_{n,l}){\tq}^{R(1)}_l(\rho_{n,l})\cdot
r_{n,k}^{n-1}\chi_k (-1)^{(n-1)\rho_{n,k}}
\times
\ny\\[-1mm] \fl &&\hs\hs
\label{mat-ele}
\left.\times\;
\frac{a_n b_n c_n}{r_{n,0}}
\left(1+\frac{(-1)^{\rho_{n,k}} d_n}{\kp_n c_n r_{n,k}}\right)
\frac{1}{\prod_{s\ne k} (r_{n,k}(-1)^{\rho_{n,k}}-r_{n,s}(-1)^{\rho_{n,s}})}
\right)\!.
\eea
The product on the right hand side of the first line comes from the change of normalization from \r{psiupsi}
to \r{mat-ele} by the factor
$\langle \tilde\Psi_{0,\bdr' }|\tilde\Psi_{0,\bdr'}\rangle/\langle
\tilde\Psi_{0,{\bf 0}}|\tilde\Psi_{0,{\bf 0}}\rangle$,
which can simply be read off from \r{norm} with $C_n$ cancelling.

We have not yet been able to perform the summation over $\bdr'_n$ in \r{mat-ele}
for general parameters of the BBS-model. However, for the homogenous Ising case \r{spIsi}, we
can show that \r{mat-ele} can be put into a fully factorized form,
although in this case special complications appear from the coincidence of the zeros
of the transfer matrix with the polynomial zeros of $B_n(\la)$.
In the next section we give the result of this summation,
which proves a formula for the matrix elements of the spin operator (form-factors)
on a finite lattice conjectured \cite{BL1,BL2} by A.~Bugrij and O.~Lisovyy.
Details of the proof which is a new result for the Ising model,
and the comparison with the notation of \cite{BL1,BL2} are relegated to the sequel article \cite{gipst2}.

\section{The homogeneous Ising model}
\label{sect6}

\subsection{Relation to the standard Ising model}\label{sect61}

In this section we restrict the parameters of the homogeneous $N=2$ BSS model to be
$a=c$, $b=d$ and $\kp=1$. So the cyclic $L$-operator \r{bazh_strog} reduces to
\be
L_k(\lm)=\lk\ba{cc}
1+\lm\,\hv_k &  \lm \,\hu_k\, (a\,-b\, \hv_k)\\
\hu_k\, (a\,-b\, \hv_k) & \lm\, a^2 + \hv_k\, b^2
\ea \rk.\ee
Let us make the special choice of the spectral parameter $\lm=b/a$ as in \r{spIsi}. Then the
$L$-operator degenerates and we get
\be \fl L_k(b/a)=(1\,+{\bf v}_k\,b/a\,)\lk\ba{cc}
1 &  {b}\, \hu_k \\a\, \hu_k & a\,b\ea \rk\,=\,
(1\,+{\bf v}_k\,b/a\,)
\lk \ba{c} 1\\ a\, \hu_k \ea \rk
\lk \ba{cc} 1\: &  b\, \hu_k\!\ea \rk.\ee
At this point the transfer matrix is
\be\fl
{\bf t}_n(b/a)=\tr\; L_1(b/a)L_2(b/a)\cdots L_n(b/a)=
\prod_{k=1}^n (1+{\bf v}_k \cdot {b}/{a})\cdot
\prod_{k=1}^n (1+\hu_{k}\hu_{k+1}\cdot a\,b).
\ee
Recall that due to the periodic boundary conditions $\hu_{n+k}\equiv\hu_{k}$.
Using
\[\fl
\exp (K_x \hu_{k}\hu_{k+1})=\cosh K_x (1\!+\hu_{k}\hu_{k+1} \tanh K_x),\;
\exp (K^*_x \hv_k)=\cosh K^*_x (1\!+\hv_k \tanh K^*_x)
\]
and writing $\:\hu_k\,=\:\sigma^z_k\,$ and $\:\hv_k\,=\:\sigma^x_k$,
it is easy to identify ${\bf t}_n(b/a)$ with the standard Ising
transfer-matrix:
\[
{\bf t}_{\rm Ising}=\exp{\lk \sum_{k=1}^n\, K^*_x\, \sigma^x_k\rk}\:
\exp{ \lk\sum_{k=1}^n\, K_x \,\sigma^z_{k}\,\sigma^z_{k+1}\rk},
\]\[
e^{-2K_y}= \tanh K^*_x=\frac{b}{a}\,,\quad\; \tanh K_x=ab\,,
\]
where $K_{x,y}$ and $K^*_{x,y}$ are coupling constants of Ising and dual Ising models
on the square lattice along the $X$ and $Y$ axes, respectively.

The eigenvectors are obtained by the method of separation of variables,
they do not depend on $\lm$.
In what follows we shall {\it not}~ fix $\,\lm=b/a\,$ and
so we consider a family of models depending on $\lm$, $a$ and $b$.
This family includes the Ising model at $\lm=b/a$ and
the eigenvectors obtained by SoV are eigenvectors of the Ising transfer matrix  too.
Let us note that this $L$-operator formulation can be extended \cite{Bugrij} to a larger family
 of Ising-like models,
giving a possibility to use SoV for finding explicit formulas for the
eigenvectors of corresponding transfer-matrices.

In \r{funceq} and \r{tNSR} from \cite{gips} we already quoted
the functional equation and the eigenvalues of the
{\it general}~ homogeneous $N=2$ BBS model. Introducing
\be \fl {\cal T}(\la^2)\:=\,1\,+b^2d^2/\kp^2\,-\lm^2(\kp^2+a^2c^2),\hs F(\lm)=
            \lk b\,- a \kp \lm\rk \lk \la c+d/\kp \rk,\ee
the equation $\,A(\qu)\la_\qu^2-C(\qu)+2i\la_\qu B(\qu)\,=\,0\,$ can be written as
\be\label{sp-eq}{\cal T}(\lms_\qu^2)\,=\,e^{{\rm i}\qu}\,F(\lms_\qu)\,+\,e^{-{\rm i}\qu}\,F(-\lms_\qu)\,.
\ee
The solution of \r{sp-eq} is given by (\ref{57}) where $\qu=\pm\frac{\pi}{n}\,m$, with
$m$ even for the R-sector $(\rho=1)$ and $m$ odd for the NS-sector $(\rho=0)$.

The equations for the amplitudes $r_{n,k}$ of the roots of the eigenvalue polynomial of the operator
$B_n(\lm)$ of equation \r{Blm} were derived in equation (A7) of \cite{gips}:\footnote{In \cite{gips}
we used $\phi_{n,k}/2$ instead of $q_{n,k}$.}
\be\label{am-eq}\fl
{\cal T}(r_{n,k}^2)^2=4 \cos^2(q_{n,k}) F(r_{n,k})F(-r_{n,k})\,,\hq \:q_{n,k}=\pi k/n, \hq k=1,2,\ldots,n-1.
\ee

In the case of the Ising model parametrization \r{spIsi} we have
$F(\lm)=b^2-a^2 \lm^2$, and \r{sp-eq} and \r{am-eq} reduce to
$\,{\cal T}(\lms_\qu^2)=2\cos (\qu)\, F(\lms_\qu)\,$ and
$\,{\cal T}(r_{n,k}^2)^2\,=\,4 \cos^2(q_{n,k})\, F^2(r_{n,k})\,$, respectively.
Then the solutions of these equations are:
\be   s_{\qu}\,=\,\sqrt{\frac{b^4-2 b^2\cos{\qu}+1}{a^4-2 a^2\cos{\qu}+1}}\,,\hq
r_{n,k}\,=\,\sqrt{\frac{b^4-2 b^2\cos{q_{n,k}}+1}{a^4-2 a^2\cos{q_{n,k}}+1}}\,, \label{solsrnk}
\ee
where the momentum $\qu$ in both NS- and R-sectors takes the values $\{0,\pi,\pm q_{n,k}\}$ with
$0<q_{n,k}<\pi$, in particular $q_{n,k}\neq 0,\:\pi$. So we have
\be\label{s-sp-eq}
r_{n,k}\,=\,s_{q_{n,k}}\,=\,s_{-q_{n,k}}\,,\quad
\lms_{0}\,=\frac{b^2-1}{a^2-1}\,,\qquad \lms_{\pi}=\frac{b^2+1}{a^2+1}\,.
\ee
The possible coincidence in the Ising case of the two quite different parameters:
the zeros $r_{n,k}$ of the polynomial $B_n(\la)$ and the zeros $s_\qu$
of transfer matrix
will create some peculiarities when in the following we set out to solve the Baxter equations.

\subsection{Solution of the Baxter equations}\label{solubax}

We have seen that in the Ising case \r{spIsi} the eigenvectors and eigenvalues of
${\bf t}_n(\la)$
decompose into two sectors $\rho=0,\,1$. With \r{spIsi} in addition we have also $\:F(\la)=F(-\la)\,$
and the Baxter equations \r{BAXL}, \r{BAXR} for left and right kernels become identical.
Omitting the superscripts $L$ and $R$ on $Q_k$ and writing
$\;\la_{n,k}=-(-1)^{\rho_{n,k}}r_{n,k}, \;\;\rho_{n,k}=0,1\:$ we obtain:
\be \fl t(\la_{n,k})\,Q_k(\rho_{n,k}) \,=\,\lk\frac{(-1)^\rho F^{n-1}(\la_{n,k})}{(\la_{n,k})^{n-1}\,\chi_k}
\:+\:(-\la_{n,k})^{n-1}\,\chi_k\:F(\la_{n,k})\:
     \rk Q_k(\rho_{n,k}+1).  \label{BaxIs}\ee
In order to solve these Baxter equations for a
given sector $\rho$, we need $t(\pm r_{n,k})$. Now from \r{tNSR} we have
\be\label{tINSR}
t(\lm)=(a^{2n}\,+(-1)^\rho)\;{\textstyle \prod_{\qu}}\:
    \lk \lm\,+\,\lm_\qu\rk,\hs \lm_\qu=(-1)^{\sigma_{\qu}}\lms_{\qu},\ee
and we see that $\,t_n(\pm r_{n,k})\,$ vanishes
if $\:\la_\qu=\mp r_{n,k} $.
When $\rho$ and $k$ have the same parity
$(-)^\rho=(-)^k$ the quasi-momenta $\qu$ which describe the spectrum $\lm_\qu$ of the transfer matrix
do not coincide with any $q_{n,k}$ and the
value of the transfer matrix $t(\pm r_{n,k})$ does not  vanish. If $\rho$ and $k$ have different
parities: $\,(-)^\rho=(-)^{k+1}\,$ this could happen. Below, when we solve the Baxter equations
we shall treat these cases separately.

Moreover, as we can see from \r{s-sp-eq}, for the Ising parameterization
the amplitudes $\lms_{q_{n,k}}=\lms_{-q_{n,k}}$ coincide. This leads to a
degeneracy of eigenvalues of ${\mathbf t}_n(\lm)$: an eigenstate with $\qu=q_{n,k}$ excited and
$\qu=-q_{n,k}$ not excited
has the same eigenvalues as the eigenstate with $-q_{n,k}$ excited and $q_{n,k}$ not excited:
both have the same factor $(\lm^2-\lms_{q_{n,k}}^2)$ in the eigenvalue polynomial.
The degeneracy of eigenvalues of the transfer-matrix can lead potentially to the following problem:
the functional relation guarantees, of course, the existence of a non-trivial solution of the
Baxter equations for
any particular eigenvalue-polynomial, but we need in our case two independent solutions.
Fortunately, due to the coincidence $\lms_{\qu=\pm q_{n,k}}=r_{n,k}$ {\it both}
sides of the Baxter equations
for the unknowns ${\tq}_k(0)$ and ${\tq}_k(1)$ become zero
and we can build a two-dimensional solutions space.

The compatibility condition following from \r{BaxIs} is
\[\fl
t(-r_{n,k}) t(r_{n,k}) =(-1)^{n-1 }
\left(\frac{(-1)^\rho\:F^{n-1} (r_{n,k})}{(r_{n,k})^{n-1}\,\chi_k}
\:+\:(-r_{n,k})^{n-1}\,\chi_k\:F(r_{n,k})\right)^2,
\]
where $t(\lm)$ is an eigenvalue from the sector $\rho$.
If $\,k\,$ is such that $\,(-1)^{k}=(-1)^{\rho+1}$, then the quasi-momentum $\qu=q_{n,k}$
belongs to the sector $\rho$
and for $r_{n,k}=\lms_{q_{n,k}}$ we have $t(-r_{n,k})t(r_{n,k})=0$. This implies
a relation not depending on a particular $t(\lm)$ and its $\rho$:
\be\label{shchi}
\chi_k^2\: r_{n,k}^{2(n-1)}=(-1)^{n+k+1}F^{n-2}(r_{n,k})\,.
\ee

In order to find the eigenvector corresponding to $t(\lm)$ from the sector $\rho$ we have
to find all the $\tq_k(\rho_{n,k})$ for all $k$ solving the Baxter equations.
We need to distinguish the following four cases with respect to the value $k$:

\medskip
\noindent (i)
$(-1)^\rho=(-1)^{k}$.
In this case both $\,t(r_{n,k})\ne 0$ and $t(-r_{n,k})\ne 0$.
So we may fix $\:{\tq}_k(0)\,=\,1\;$ and,
using \r{shchi} we obtain
\[
{\tq}_k(1)\:=\:
\frac{(-1)^{n-1} t(-r_{n,k})}{2 \chi_k r_{n,k}^{n-1}\:F(r_{n,k})}\,.
\]
The other three cases correspond to $(-1)^\rho=(-1)^{k-1}$ so that
the big brackets on the right-hand sides of the Baxter equations are zero due to (\ref{shchi}).

\medskip
\noindent (ii)
$t(r_{n,k})\ne 0,\ t(-r_{n,k})=0$:
The eigenvalue polynomial $t(\lm)$ contains the factor $(\lm+r_{n,k})^2$ (both momenta
$\qu=\pm q_{n,k}$ are not excited, i.e. both not in the spectrum) and
we have\\[-9mm]
\[
{\tq}_k(0)\:=\:1\,,\qquad
{\tq}_k(1)\:=\:0\,.
\]

\medskip
\noindent (iii)
{$t(r_{n,k})=0,\ t(-r_{n,k})\ne 0$:}
$t(\lm)$ contains the factor $(\lm-r_{n,k})^2$ (both momenta
$\qu=\pm q_{n,k}$ are excited) and
we have\footnote{In this case we can't normalize
$\;\tq_k(0)\:=\:1$.}
\[
{\tq}_k(0)\:=\:0\,,\qquad
{\tq}_k(1)\:=\:1\,.
\]

\medskip
\noindent (iv)
{both $\:t(r_{n,k})=0\,$ and $\:t(-r_{n,k})=0$.}
This case happens when only one of the two quasi-momenta $\qu=\pm q_{n,k}$ is excited, so that $t(\lm)$
contains a factor $(\lm^2-r_{n,k}^2)$.
As we explained above, this eigenvalue is degenerate.
In this case both hand sides of both Baxter equations are zero.
In principle, we can choose any two independent solutions for Baxter equations,
but, in general, they will not give us eigenvectors of operator of translation ${\bf T}$.
Since our construction of the eigenvectors  \r{EV} is obviously non-invariant with respect
to translations
it is unclear how using the direct action of ${\bf T}$ one can choose
solutions of the Baxter equations which give such eigenvectors.
We take another way:
In order to obtain eigenvectors of transfer matrix which are eigenvectors
also of the translation operator,
we shall first lift the degeneracy by starting with parameters such that $ad-bc=\eta$
(keeping $\kp=1$) with $\eta$ small but finite. At the end we shall take the limit $\eta\rightarrow 0$.
Observe that for $\eta\neq 0$ \r{BEq} and \r{LBaxter} are different. From \r{BEq} we get
\[\fl {\tq}^{\rm R}_k(0)=1,\hq\!\!
{\tq}^{\rm R}_k(1)=\frac{t(-r_{n,k})}{(-1)^\rho/(\chi_k\,(r_{n,k})^{n-1})\,F^{n-1}(r_{n,k})
+(-1)^{n-1} \chi_k (r_{n,k})^{n-1}\:F(-r_{n,k})}\,.
\]
When $\eta\to 0$, both numerator and denominator of ${\tq}^{\rm R}_k(1)$ tend to $0$.
Let us find their leading (in fact, linear) terms in $\eta\to 0$.
For the denominator we have for $\:\eta\to 0$
\[\fl
\frac{(-1)^\rho\,F^{n-1}(r_{n,k})}{\chi_k\,(r_{n,k})^{n-1}}\:
+(-1)^{n-1}\chi_k (r_{n,k})^{n-1}\,F(-r_{n,k})
\sim \eta\: (-1)^{n+1} n  r_{n,k}\chi_k\,(r_{n,k})^{n-1}\,,
\]
where we used the relation (valid for arbitrary $\eta$)
\[
F^{n-2}(r_{n,k})\, F^{n-2}(-r_{n,k})\,=\,\chi_k^4 \;r_{n,k}^{4(n-1)}\,.
\]
Let us write the transfer-matrix eigenvalue polynomial \r{tINSR} as
\[
t(\lm)={t}_{\check \qu_k}(\lm)\;(\lm-\lms_{\check\qu_k})(\lm+\lms_{-{\check\qu_k}})\,,
\]
where ${\check\qu_k}=(-1)^{\sigma_{q_{n,k}+1}}q_{n,k}$ is the value of the excited quasi-momentum,
so that
\[
t(-r_{n,k})={t}_{\check \qu_k}(-r_{n,k})\:(r_{n,k}-\lms_{-{\check\qu_k}})(r_{n,k}+\lms_{{\check\qu_k}})\,.
\]
Using $\;A(\qu)(\la\,-\lms_{-\qu})(\la\,+\lms_{\qu})= A(\qu)\,\lm^2\,-\,C(\qu)\,-2{\rm i}\,B(\qu) \lm$,
$\;B(\qu)\,=\eta\, \sin \qu\;$ and
$A(\check \qu_k)\,r_{n,k}^2\,-\,C(\check \qu_k)\sim o(\eta)$ (due to
\r{C} and \r{57} at $\eta\to 0$) we get
\[\fl
t(-r_{n,k})\sim -{ t}_{{\check \qu_k}}(-r_{n,k})2{\rm i}\,B({\check \qu_k})\:r_{n,k} /A({\check \qu_k})
=- {t}_{\check \qu_k}(-r_{n,k})2{\rm i}\sin (\check \qu_k)\: r_{n,k}\eta /A(\check \qu_k)\]
at $\eta\to 0\,,\;$ where $A(\qu)=a^4-2a^2\cos \qu\,+1$.
Finally,
\[{\tq}^{\rm R}_k(0)=1\,,\qquad
{\tq}^{\rm R}_k(1)=
\frac{(-1)^{n+\sigma_{q_{n,k}}+1}  2{\rm i}\sin (q_{n,k})\:{t}_{{\check \qu_k}}(-r_{n,k})}
{ n \chi_k\,r_{n,k}^{n-1}\: A(q_{n,k})}
\,.
\]
Using a similar limiting procedure for the Baxter equation \r{LBaxter} we get
\[\hq{\tq}^{\rm L}_k(0)=1\,,\hs
{\tq}^{\rm L}_k(1)=
-{\tq}^{\rm R}_k(1)\,.\]
Although the Ising Baxter equations were the same for $Q^L$ and $Q^R$,
in this case, since we take the limit of the general non-hermitian BBS case,
$Q^L_k(1)$ and $Q^R_k(1)$ come out different.

\subsection{Norms and factorized matrix elements for the homogeneous Ising model}

In this Subsection we state a new factorized formula for the matrix element of the spin operator
in the Ising model on a finite lattice.
Let $|\Phi_0\rangle$ and $|\Phi_1\rangle$ be two eigenvectors of the periodic Ising model from
the sectors $\rho=0$ and $\rho=1$, respectively.
First let us consider the case when these states do not contain fermionic excitations with
momenta $\qu=q_{n,k}$ and $\qu=-q_{n,k}$ simultaneously, so that the eigenvalue polynomials
do not contain a factor $(\lm-r_{n,k})^2$.
We introduce the subset ${\cal D}\subset\{1,2,\ldots,n-1\}$ of
indices $k$ for which the eigenvalue polynomials of $|\Phi_0\rangle$ and
$|\Phi_1\rangle$ contain the factor $(\lm^2-r_{n,k}^2)$, i.e. for which we have the
case (iv) of~ Subsection \ref{solubax}.
Denote by $|{\cal D}|$ the size of the set ${\cal D}$.

Let $\delta=1$ if ${\sigma_0}={\sigma_\pi}$ and
$\delta=0$ otherwise. We also use the short notations: $\lm_0=(-1)^{\sigma_0}\lms_0$,
$\lm_\pi=(-1)^{\sigma_\pi}\lms_\pi$ and $\hat q_k=(-1)^{\sig_{q_{n,k}}+k}q_{n,k}$.
Now the matrix element is given by the factorized formula:
\bea\fl
\lefteqn{\frac{\langle \Phi_0|\bu_n|\Phi_1\rangle\;\langle \Phi_1|\bu_n|\Phi_0\rangle}
{\langle \tilde\Psi_{0,{\bf 0}}|\tilde\Psi_{0,{\bf 0}}\rangle^2}\;=\;
(\lm_\pi^2-\lm_0^2)^{(|{\cal D}|-\delta)/2}\:
(\lm_0+\lm_\pi)^{\delta}\;\times}\ny\\
\fl \hs\hs\hs &\times&\prod_{k \in {\cal D}}
\! \frac{2r_{n,k} }{ (r_{n,k}+\lm_0)(r_{n,k}+\lm_\pi) }
\!\prod_{k<m \atop k,m \in {\cal D}}\frac{r_{n,k}\!+r_{n,m}}{r_{n,k}\!-r_{n,m}}
\!\prod_{k<m \atop k,m \in {\cal D}} \frac{\sin\frac{1}{2}(\hat q_k-\hat q_m)}
{\sin\frac{1}{2}(\hat q_k+\hat q_m)}\,.
\label{ME} \eea
The proof of \r{ME} is given in our sequel article \cite{gipst2}.

In order to compare \r{ME} to the results obtained by  A.~Bugrij and O.~Lisovyy
\cite{BL1,BL2} we have to calculate instead of \r{ME} the ratio
\be\label{prq}\frac{\langle \Phi_0|\bu_n|\Phi_1\rangle \langle \Phi_1|\bu_n|\Phi_0\rangle }
 {\langle \Phi_0|\Phi_0\rangle\langle \Phi_1|\Phi_1\rangle}\,,
\ee
which has the advantage to be independent of the particular normalization of
$|\Phi_0\rangle$ and $|\Phi_1\rangle$.
To do this we have to divide \r{ME} by
\be\label{prq1}\langle \Phi_0|\Phi_0\rangle
\langle \Phi_1|\Phi_1\rangle/
\langle \tilde\Psi_{0,{\bf 0}}|\tilde\Psi_{0,{\bf 0}}\rangle^2\,.
\ee

The expression for \r{prq1} was obtained in  (\ref{normP}) for generic parameters.
In the case of the Ising model the norms of the degenerated states
cannot be directly obtained from \r{normP}, because the numerator and denominator contain
zeros. When degenerated states appear,
that is when ${\cal D}$ is not empty, we have to be careful in
using the formula (\ref{normP}) for the norms and use l'H{\^o}pital's rule.
In order to implement l'H{\^o}pital's rule we proceed similarly
to what we did in case (iv) of Section 6.2 for finding the solutions of Baxter equations:
we go off the Ising point taking $\;ad-bc\,=\eta\;$ to be small finite and keep terms linear in $\eta$.
Let ${\check\qu_l}=(-1)^{\sigma_{q_{n,l}+1}}q_{n,l}$ for $l\in {\cal D}$.
Then the coefficient of $\lm$ in
\be\label{ABCqu}
A(\check\qu_l)(\lm\,-\lms_{{\check\qu_l}})(\lm\,+\lms_{-{\check\qu_l}})=
A(\check\qu_l)\,\lm^2\,-\,C(\check\qu_l)\,+2{\rm i}\,B(\check\qu_l) \lm
\ee
gives $-\lms_{\check\qu_l}+\lms_{-\check\qu_l}=2{\rm i}\,B(\check\qu_l)/A(\check\qu_l)$.
Due to $A(\check \qu_l)\,r_{n,l}^2\,-\,C(\check \qu_l)\sim o(\eta)$ (when $\eta\to 0$)
the formula \r{ABCqu} at $\lm=r_{n,l}$
gives
$(r_{n,l}-\lms_{\check\qu_l})(\lm\,+\lms_{-\check\qu_l})\sim 2 {\rm i}\,B(\check\qu_l)
r_{n,l}/A(\check\qu_l)$.
Therefore
\[
\frac{-\lms_{\check\qu_l}+\lms_{-\check\qu_l}}{(r_{n,l}-\lms_{\check\qu_l})(r_{n,l}+\lms_{-\check\qu_l})}
\to \frac{1}{r_{n,l}}\hs\mbox{for}\hx \eta\to 0
\]
for the corresponding factor in the formula (\ref{normP}) for norm. Note that this result
is independent of which of the two quasi-momenta $\:q_{n,l}$ or $-q_{n,l}$ is excited.
Now we have to take the product of one term \r{normP} for R with another for NS.
Since the terms at $s_\pi$ appear only in NS (R) for $n$ odd (even),
we get two slightly different formulas for these cases:

For $n$ odd we have
\[\fl
\frac{\langle \Phi_0|\Phi_0\rangle\;\langle \Phi_1|\Phi_1\rangle}
{\langle \tilde\Psi_{0,{\bf 0}}|\tilde\Psi_{0,{\bf 0}}\rangle^2}
\;=\;2^{|{\cal D}|} \prod_{k=1}^{n-1} (2r_{n,k}) \cdot \frac{\prod_{k{\rm -odd}}(\lm_\pi\pm r_{n,k}) }
{\prod_{k \rm{-even}} (\lm_\pi+r_{n,k})}\cdot
\frac{\prod_{k \rm{-even}} (\lm_0\pm r_{n,k}) }
{\prod_{k \rm{-odd}} (\lm_0+r_{n,k})}\times
\]
\be\fl\times
\frac{ \prod_{k<l, k,l{\rm -odd}} \Bigl((r_{n,k}+r_{n,l}) (\pm r_{n,k}\pm r_{n,l})\Bigr)
\prod_{k<l, k,l{\rm -even}} \Bigl((r_{n,k}+r_{n,l}) (\pm r_{n,k}\pm r_{n,l})\Bigr)}
{\prod_{k{\rm -odd},l{\rm -even}}  \Bigl((\pm r_{n,k}+r_{n,l}) (r_{n,k}\pm r_{n,l})
\Bigr)}\,,\label{normodd}
\ee
and for $n$ even
\[\fl\frac{\langle \Phi_0|\Phi_0\rangle\;\langle \Phi_1|\Phi_1\rangle}
{\langle \tilde\Psi_{0,{\bf 0}}|\tilde\Psi_{0,{\bf 0}}\rangle^2}
\;=\;
2^{|{\cal D}|} \prod_{k=1}^{n-1} (2r_{n,k}) \cdot (\lm_0+\lm_\pi)
\frac{\prod_{k \rm{-even}} (\lm_0\pm r_{n,k}) (\lm_\pi\pm r_{n,k})}
{\prod_{k \rm{-odd}} (\lm_0+r_{n,k}) (\lm_\pi+r_{n,k})}\times
\]\be\fl\times
\frac{ \prod_{k<l, k,l{\rm -odd}} \Bigl((r_{n,k}+r_{n,l}) (\pm r_{n,k}\pm r_{n,l})\Bigr)
\prod_{k<l, k,l{\rm -even}} \Bigl((r_{n,k}+r_{n,l}) (\pm r_{n,k}\pm r_{n,l})\Bigr)}
{\prod_{k{\rm -odd},l{\rm -even}}  \Bigl((\pm r_{n,k}+r_{n,l}) (r_{n,k}\pm r_{n,l})
\Bigr)}\,, \label{normeven}
\ee
where the sign at $\pm r_{n,m}$ is fixed `$-$' if $m\in {\cal {D}}$
and `$+$' otherwise.

The final result \r{prq} will be given in terms of $\lm_0$, $\lm_\pi$ and $r_{n,k}$, $k=1,2,\ldots,n-1$.
Now if the eigenvalue polynomials contain the factors
$(\lm-r_{n,l})^2$ instead of $(\lm+r_{n,l})^2$ for some $l$
we have to replace $r_{n,l}\to -r_{n,l}$
in the final formula for the matrix element \r{prq} for all such $l$.

Let us finish this section by giving as an example the matrix element for the $3$-site chain between
$|\Phi_0\rangle=|\pi,\pi/3\rangle_{\rm\, NS}$ and
$|\Phi_1\rangle=|0\rangle_{\rm\, R}$. Using \r{normodd} we  have
\[
\frac{\langle \Phi_0|\,\Phi_0\rangle\;\langle \Phi_1|\,\Phi_1\rangle}
{\langle \tilde\Psi_{0,{\bf 0}}|\,\tilde\Psi_{0,{\bf 0}}\rangle^2}
\;=\;
-\frac{8 r_{31}r_{32}  (r_{31}+\lms_\pi)(r_{32}-\lms_0)}
{(r_{32}^2-r_{31}^2)(r_{31}-\lms_0)(r_{32}-\lms_\pi)}\,.
\]
Since in this case ${\cal D}=\{1\}$, formula \r{ME} gives
\[\frac{\langle \Phi_0|\,\bu_3|\Phi_1\rangle\;\langle \Phi_1|\,\bu_3|\Phi_0\rangle}
{\langle \tilde\Psi_{0,{\bf 0}}|\,\tilde\Psi_{0,{\bf 0}}\rangle^2}\;=\;
-\frac{2r_{31} (\lms_0+\lms_\pi)}{(r_{31}-\lms_0)(r_{31}-\lms_\pi)}\,.
\]
Finally,\\[-9mm]
\[
\frac{\langle \Phi_0|\bu_3|\Phi_1\rangle \langle \Phi_1|\bu_3|\Phi_0\rangle}
{\langle \Phi_0|\Phi_0\rangle\langle \Phi_1|\Phi_1\rangle}=
\frac{ (\lms_0+\lms_\pi)(\lms_{2\pi/3}^2-\lms_{\pi/3}^2)(\lms_{2\pi/3}-\lms_\pi)}
{4\lms_{2\pi/3}(\lms_{\pi/3}^2-\lms_\pi^2)(\lms_{2\pi/3}-\lms_0)}\,,
\]
where due to \r{solsrnk}
\[
\lms_{\pi/3}^2=r_{31}^2=\frac{b^4-b^2+1}{a^4-a^2+1}\,,\qquad
\lms_{2\pi/3}^2=r_{32}^2=\frac{b^4+b^2+1}{a^4+a^2+1}\,.
\]

\section{Conclusions}
In this paper we continue our calculation of state vectors and matrix elements of the finite-size
inhomogenous
$\mathbb{Z}_N$-Baxter-Bazhanov-Stroganov
lattice spin model, using the method of ~Separation of Variables. In a previous paper \cite{gips} we
gave the right eigenvectors for the auxiliary system and the Baxter equations which determine the
periodic boundary condition BBS-eigenvectors. Here we complete this work by calculating explicit
formulae for the corresponding left eigenvectors and norms and matrix elements of a single operator.
A main result is the expression of Theorem 1 for the norm of the state vectors of the auxiliary
system: the multiple summation over the intermediate states is performed,
so that the norm is put into a factorized
form. For $N=2$ the Baxter equations are solved explicitly, and also the norm of the periodic
model is put into a compact factorized form. Since it is an open task to perform
the intermediate spin summations
for matrix elements of the inhomogenous periodic model, in the last Section
we proceeded with the homogenous Ising
model parameters. There we manage to perform the summations and we present
a new factorized formula for the matrix elements of spin operator,
delegating the somewhat lengthy derivation and comparison with the corresponding formula
in \cite{BL1,BL2} to our sequel paper \cite{gipst2}.

\section*{Acknowledgements}
This work has been supported by the Heisenberg-Landau program HLP-2007.
SP was supported by the RFBR grant 05-01-01086 and grant for support of scientific schools
NSh-8065.2006.2, NI and VS  were supported by INTAS grant 05-1000008-7865 and Ukrainian DFFD grant.
GvG is grateful to the Department
of Physics of Complex Systems of the Weizmann Institute for kind hospitality.

\setcounter{section}{0}
\setcounter{subsection}{0}

\renewcommand{\thesection}{\Alph{section}}

\section*{Appendix: The main summation formula}

Consider a degree $n$ polynomial $f(t)=f_n\,t^n+\ldots+f_0$ and its interpolation formula through
$n+1$ arbitrary not coinciding points $\:c_1,\ldots,c_{n+1}$:
\be    f(t)\;=\;\sum_{k=1}^{n+1}\;f(c_k)\;\prod_{s\neq k}^{n+1}\;
                    \frac{t-c_s}{c_k-c_s}\,,\qquad
 f_n\;=\;\sum_{k=1}^{n+1}\;\frac{f(c_k)}{\prod_{s\neq k}^{n+1}(c_k-c_s)}\,.
 \label{sumru}\ee
We shall use \r{sumru} for $f_n=0$, i.e. for polynomials $f(t)$ of degree less than $n$,
obtaining a sum rule valid for an arbitrary choice of the parameters $\lb c_1,\ldots,c_{n+1}\rb$:
\be\sum_{k=1}^{n+1}\;\frac{f(c_k)}{\prod_{s\neq k}^{n+1}(c_k-c_s)}\;=\;0\,.\label{suru}\ee
Now take an arbitrary polynomial $f(t)$ of degree less than $n+N$ and choose
$n+N$ points $\;(c_1,\,\ldots,\,c_{n+N})\:=\:(b_1,\ldots,b_n,\,r,\om
r,\ldots,\om^{N-1}r)\,.\:$  Then the sum rule corresponding to \r{suru} is\\[-8mm]
\be
 \sum_{\rho=0}^{N-1}\:\frac{\om^\rho r\, f(\om^\rho r)}
  {\prod_{s=1}^n\:(\om^\rho r-b_s)}= Nr^N
 \sum_{k=1}^{n}\;\frac{f(b_k)}{(r^N\,-\,b_k^N)\:
 \prod_{s=1\atop s\neq k}^{n}\;(b_k\,-\,b_s)}
\label{sumrho}
\ee
since $\;\; \prod_{i=0,\:i\neq
\rho}^{N-1}(\om^\rho r\,-\om^i r)\;=\;N\om^{-\rho}\:r^{N-1}. $\\[-3mm]

The goal of this Appendix is to prove equation \r{SnnInt} of the main text,
i.e. we shall show by induction that
 for the sets
$\mathbf{a}^{(n)}=\{a_1,\ldots,a_{n-2}\}=\{r_1\om^{-\rho_1},\ldots,r_{n-2}\om^{-\rho_{n-2}}\}$ and
$\mathbf{b}^{(n)}=\{b_1,\ldots,b_{n-1}\}$
the ${\mathbb Z}_N$-symmetrical sum over all phases of the parameters $a_l$ in\\[-4mm]
\be S_n(\mathbf{a}^{(n)},\mathbf{b}^{(n)})\;=\!\!\sum_{\rho_1,\ldots,\rho_{n-2}\in{(\mathbb Z}_N)^{n-2}}
\;\prod_{l=1}^{n-2}\lk
a_l\; \frac{\prod_{j=l+1}^{n-2}\lk
a_l\:-\:a_j\rk}{\prod_{k=1}^{n-1}\lk a_l\:-\:b_k\rk}\rk\label{Sn}
\ee
can be performed explicitly, so that we obtain \r{Sn} in the {\it factorized} form
\be
 S_n(\mathbf{a}^{(n)},\mathbf{b}^{(n)})\;=\;N^{n-2}
\prod_{l=1}^{n-2}a^N_l\prod_{k<k'}^{n-1}\frac{b_k^N-b_{k'}^N}{b_k-b_{k'}}\;\:
   \frac{\prod_{l<l'}^{n-2}(a_l^N-a_{l'}^N)}
   {\prod_{l=1}^{n-1}\prod_{l'=1}^{n-2}(b_l^N-a_{l'}^N)}.\label{Snn}
\ee

This assertion is correct for $n=3$, as can be seen by explicit summation, using
$ \sum_{\rho\in\ZN}\:b/(r\om^{-\rho}-b)\,=\,N\,b^N/(r^N\,-b^N) $:
\bea \fl S_3(a_1;b_1,b_2)&=&\sum_{\rho_1}\;\frac{a_1}{(a_1-b_1)(a_1-b_2)}\;=\;\sum_{\rho_1}\frac{1}{b_1-b_2}
       \lk\frac{b_2}{b_2-{r_{1}}\om^{-\rho_1}}-\frac{b_1}{b_1-{r_{1}}\om^{-\rho_1}}\rk\ny\\
       \fl&=&\frac{N}{b_1-b_2}\lk\frac{b_2^N}{b_2^N-{a_1^N}}-\frac{b_1^N}{b_1^N-{a_1^N}}\rk\,
       =\frac{N\;a_1^N\;(b_1^N-b_2^N)}{(b_1-b_2)\;(b_1^N-a_1^N)(b_2^N-a_1^N)}\,.\ny
       \eea

For general $n$ it will be useful to use the induction assumption \r{Sn} also in the form
\beq  S_n(\mathbf{a}^{(n)},\mathbf{b}^{(n)})\;=\;\TS_n(\mathbf{a}^{(n)})\;
  \prod_{k<k'}^{n-1}\frac{b_k^N-b_{k'}^N}{b_k-b_{k'}}\;\:
   \frac{1}{\prod_{l=1}^{n-1}\prod_{l'=1}^{n-2}(b_l^N-a_{l'}^N)}\eeq
with \\[-11mm]
\be
\TS_n(\mathbf{a}^{(n)})\;=\;N^{n-2}
 \lk\prod_{l=1}^{n-2}a^N_l\rk\;\prod_{l<l'}^{n-2}(a_l^N-a_{l'}^N)\,.\label{stilde}
 \ee

In proceeding with the induction proof of \r{Snn} we start separating terms involving the last summation
(over
$\rho_{n-2}$) in \r{Sn} and perform this sum using \r{sumrho} with
$\:f(t)=\prod_{m=1}^{n-3}\,(a_m-t)$:
\bea\fl
S_n&=&\sum_{\rho_1,\ldots,\rho_{n-3}}\lk\prod_{l=1}^{n-3}
a_l\:\frac{\prod_{j=l+1}^{n-3}(a_l-a_j)}{\prod_{m=1}^{n-1}(a_l-b_m)}\rk\;
\sum_{\rho_{n-2}}a_{n-2}\frac{\prod_{m=1}^{n-3}(a_m-a_{n-2})}{\prod_{s=1}^{n-1}(a_{n-2}-b_s)}\ny\\
\fl &=&\sum_{\rho_1,\ldots,\rho_{n-3}}\lk\prod_{l=1}^{n-3}
a_l\:\frac{\prod_{j=l+1}^{n-3}(a_l-a_j)}{\prod_{m=1}^{n-1}(a_l-b_m)}\rk\;N\,a_{n-2}^N\sum_{k=1}^{n-1}
\frac{\prod_{m=1}^{n-3}(a_m-b_k)}{(a_{n-2}^N-b_k^N)\prod_{s=1\atop
s\neq k}^{n-1}(b_k-b_s)}\,.\ny
\eea
The first summation in the last
equation involves also the phases of $a_m$ in the term
$\prod_{m=1}^{n-3}(a_m-b_k)$ of the last numerator. However, this
term cancels just the term $m=k$ in the denominator of the first
bracket. So, comparing to \r{Sn} for $n\rightarrow n-1$ we
can write
\be
 S_n(\mathbf{a}^{(n)},\mathbf{b}^{(n)})\;=\;N\,a_{n-2}^N\sum_{k=1}^{n-1}\;
 \frac{S_{n-1}(\mathbf{a}^{(n)}_{n-2},
 {\mathbf{b}}_k^{(n)})}{(a_{n-2}^N-b_k^N)
\prod_{s=1\atop
s\neq k}^{n-1}(b_k-b_s)}\,, \label{recs}
\ee
where $\:{\mathbf{b}}_k^{(n)}$ is the
$n\!-2$ component vector resulting from $\:{\mathbf{b}}^{(n)}$
omitting the component $b_k$. Similarly, $\:\mathbf{a}^{(n)}_{n-2}$ is the
  $n\!-3$-component vector resulting from
$\:\mathbf{a}^{(n)}$ by omitting the component $a_{n-2}$.
Now we can use the induction assumption \r{Snn}
to insert here
$S_{n-1}$:
\bea\fl
\lefteqn{S_n(\mathbf{a}^{(n)},\mathbf{b}^{(n)})\;=\;N\,a_{n-2}^N\sum_{k=1}^{n-1}\;
 \frac{1}{(a_{n-2}^N-b_k^N)\prod_{s=1\atop
s\neq k}^{n-1}(b_k-b_s)}\:\times}\ny\\
&&\hspace*{-1cm}\times\:N^{n-3}
 \lk\prod_{l=1}^{n-3}a^N_l\rk\prod_{m<m'\atop m,m'\neq k}^{n-1}\frac{b_m^N-b_{m'}^N}{b_m-b_{m'}}\;\:
   \frac{\prod_{l<l'}^{n-3}(a_l^N-a_{l'}^N)}
   {\prod_{l=1\atop l\neq k}^{n-1}\prod_{l'=1}^{n-3}(b_l^N-a_{l'}^N)}\,. \eea
It is more convenient to write \r{recs} in terms of $\:\TS_n(\mathbf{a}^{(n)})\,,$ equation \r{stilde},
since then the $\,\mathbf{b}$-dependence is explicit:
\bea \fl\lefteqn{\TS_n(\mathbf{a}^{(n)})\;\prod_{k'<k''}^{n-1}\;
\frac{b_{k'}^N-b_{k''}^N}{b_{k'}-b_{k''}}\;
\frac{1}{\prod_{l=1}^{n-1}\prod_{l'=1}^{n-2}(b_l^N-a_{l'}^N)}\;=
  \;N\: a_{n-2}^N\:\TS_{n-1}(\mathbf{a}^{(n)}_{n-2})\,\times}\ny\\
\fl \hs &\times& \sum_{k=1}^{n-1}\:\prod_{k'<k''\atop k',k''\neq k}^{n-1}
\;\frac{b_{k'}^N-b_{k''}^N}{b_{k'}-b_{k''}}\;\:
  \frac{1}{\prod_{l=1\atop l\neq k}^{n-1}\prod_{l'=1}^{n-3}(b_l^N-a_{l'}^N)}
\;\:\frac{1}{(a_{n-2}^N-b_k^N)\prod_{s=1\atop s\neq k}^{n-1}(b_k-b_s)}\,.\ny
\eea
Apart from a sign, the terms linear in the $b_{k'}$ cancel since
\beq
\fl\frac{\prod_{k'<k''}^{n-1}(b_{k'}-b_{k''})}{\prod_{k'<k''\atop k',k''\neq k}^{n-1}(b_{k'}-b_{k''})}
\;=\;\prod_{k'=k+1}^{n-1}(b_k-b_{k'})\prod_{k'=1}^{k-1}(b_{k'}-b_k)\;
   =\;(-1)^{k-1}\prod_{s=1,s\neq k}^{n-1}\,(b_k\,-b_s)\,.
   \eeq
Using the same formula for the $\;b_{k'}^N$-terms and simplifying
 the $\:b_l^N-a_{l'}^N$-terms, we get
\beq
\fl \TS_n(\mathbf{a}^{(n)})\;=\;-N\: a_{n-1}^N\:\TS_{n-1}(\mathbf{a}^{(n)}_{n-2})\;\sum_{k=1}^{n-1}\;
\frac{\prod_{l=1}^{n-3}(b_k^N-a_l^N)\;\:\prod_{l\neq k}^{n-1}(b_l^N-a_{n-2}^N)}
    {\prod_{s=1,s\neq k}^{n-1}(b_k^N-b_s^N)}\,.
    \eeq
Here the sum can be calculated using the identity
\beq  \sum_{k=1}^{n-1}\;\:\prod_{l=1}^{n-3}\lk b_k^N-a_l^N\rk
 \;\prod_{s=1\atop s\neq k}^{n-1}\: \frac{b_s^N-a_{n-2}^N}{b_k^N-b_s^N}
  \;=\;-\,\prod_{l=1}^{n-3}\lk a_l^N-a_{n-2}^N\rk \,, \eeq
so that\\[-11mm]
\beq \TS_n(\mathbf{a}^{(n)})\;=\;N\: a_{n-2}^N\:\TS_{n-1}(\mathbf{a}^{(n)}_{n-2})
         \;\prod_{l=1}^{n-3}\lk a_l^N-a_{n-2}^N\rk
         \eeq
which confirms \r{stilde} and so also \r{Snn}.$\hfill\Box$

\subsection*{References}
\bibliographystyle{amsplain}

\end{document}